\documentclass[fleqn,usenatbib]{mnras}

\usepackage{newtxtext,newtxmath}

\usepackage[T1]{fontenc}

\DeclareRobustCommand{\VAN}[3]{#2}
\let\VANthebibliography\thebibliography
\def\thebibliography{\DeclareRobustCommand{\VAN}[3]{##3}\VANthebibliography}

\usepackage{graphicx}	
\usepackage{amsmath}	
\usepackage{xcolor}	    
\usepackage{orcidlink}

\newcommand{\lya}{Lyman-$\alpha$}
\newcommand{\nbody}{$N$-body}
\newcommand{\pyccray}{\texttt{pyC$^2$Ray}}

\newcommand{\pkdgrav}{\texttt{Pkdgrav3}}

\newcommand{\modelSA}{{\tt Late\_Reionization}}
\newcommand{\modelUni}{{\tt Uniform\_Reionization}}

\newcommand{\RefereeReport}[1]{\textcolor{black}{#1}}
\newcommand{\RefereeReportTwo}[1]{\textcolor{black}{#1}}

\newcommand{\addorcid}[1]{\orcidlink{\csname orcidauthor#1\endcsname}}

\title[Detecting neutral islands with IGM tomography]{Mapping neutral islands during end stages of reionization with photometric intergalactic medium tomography}

\author[S. K. Giri et al.]{
Sambit K. Giri\addorcid{SG},$^{1,2}$\thanks{E-mail: \href{mailto:s.k.giri@rug.nl}{s.k.giri@rug.nl}}
Koki Kakiichi\addorcid{KK},$^{ 3,4}$\thanks{E-mail: \href{mailto:koki.kakiichi@nbi.ku.dk}{koki.kakiichi@nbi.ku.dk}}
Michele Bianco\addorcid{MB},$^{5}$
and P. Daniel Meerburg\addorcid{DM}$^{1}$
\\
$^{1}$Van Swinderen Institute for Particle Physics and Gravity, University of Groningen, Nijenborgh 3, 9747 AG Groningen, The Netherlands\\
$^{2}$Nordita, KTH Royal Institute of Technology and Stockholm University, Hannes Alf\'vens v\"ag 12, SE-106 91 Stockholm, Sweden\\
$^{3}$Cosmic Dawn Center (DAWN), Denmark \\
$^{4}$Niels Bohr Institute, University of Copenhagen, Jagtvej 128, DK-2200 Copenhagen N, Denmark\\
$^5$Institute for Particle Physics and Astrophysics, ETH Zurich, Wolfgang-Pauli-Str 27, 8093 Zurich, Switzerland
}

\date{Accepted 2025 November 04. Received 2025 November 04; in original form 2025 May 19, NORDITA-2025-003}

\pubyear{\the\year{}}

\begin{document}
\label{firstpage}
\pagerange{\pageref{firstpage}--\pageref{lastpage}}
\maketitle

\begin{abstract}
During the epoch of reionization (EoR), the first generation of luminous sources in our Universe emitted ionizing photons that almost completely ionized the gas in the intergalactic medium (IGM). The growth of ionized bubbles and the persistence of neutral islands within the IGM hold vital clues to understanding the morphology and timeline of cosmic reionization. We explore the potential of photometric IGM tomography using deep narrow-band (NB) imaging to observe the \lya{} forest transmission in background galaxies with the Subaru/Hyper-Suprime Cam (HSC). Based on our simulations, we find that the currently available NB filter is suitable for mapping the IGM at $z\simeq 5.7$, corresponding to the late stages of reionization. Our findings indicate that over $\sim$500 background galaxies are needed to accurately reconstruct the IGM at scales greater than 200 Mpc, achieving more than \RefereeReport{a} 40 per cent correlation with the true distribution. This technique can help detect \RefereeReport{the} final remaining neutral islands that span more than 20 Mpc lengths. Using the superpixel method built to identify physical patterns in noisy image data, we find that the neutral island size distribution can be recovered with an accuracy of $\sim$0.3 dex.  Furthermore, we demonstrate that these reconstructed maps are correlated with the galaxy distribution and anti-correlated with the cosmological 21-cm signal from neutral hydrogen in the IGM. Lastly, we find that these reconstructed maps are anti-correlated with the patchy optical depth to the cosmic microwave background. As such, multiple measurements can be employed for \RefereeReport{the} confirmed detection of neutral islands during the end stages.
\end{abstract}




\section{Introduction}

When the first stars and galaxies formed, they emitted energetic photons into the intergalactic medium (IGM), heating and ionizing the surrounding gas. This transformative period, known as the epoch of reionization (EoR), holds essential clues about the early Universe's physical processes and the nature of the first luminous sources \citep{barkana2001beginning,furlanetto2006cosmology, dayal2018early}. The growth and merger of ionized bubbles during this time reveal insights into the clustering of ionizing sources and the efficiency of reionization \citep[e.g.,][]{furlanetto2006characteristic,zahn2007simulations, friedrich2011topology, kakiichi2017recovering, giri2018bubble, elbers2019persistent, zackrisson2020bubble, giri2021measuring}.

As these ionized bubbles expand around sources of ionizing radiation, they eventually merge, leading to the full reionization of the IGM. Current observations suggest that reionization has concluded around redshift $z \sim 5.3$ \citep{kulkarni2019large, bosman2022hydrogen}. In this study, we focus on the late stages of reionization ($z \lesssim 6$), when most of the IGM is ionized, but a few neutral islands remain \citep[e.g.,][]{giri2019neutral, keating2020long, nasir2020observing}. Though relatively scarce, these neutral regions leave a significant imprint on high-redshift observational data due to their large sizes \citep{giri2019neutral, giri2024end}. Mapping these residual neutral islands is crucial for understanding both the properties of the ionizing sources and the IGM, such as the ionizing background, mean free path of ionizing photons and clumpiness of matter distribution \citep{keating2020long,bosman2024measurement,davies2024predicament,davies2024constraints,cain2025new}. In this study, our objective is to explore and probe these remaining neutral regions.

Several observational techniques have been developed to probe the ionization state of the IGM during the EoR, with one of the most promising methods being the measurement of the 21-cm signal from neutral hydrogen \citep[see e.g.,][for a review]{pritchard201221}. This technique relies on the hyperfine transition of neutral hydrogen atoms to trace the distribution of neutral gas across vast regions of space. The 21-cm signal is a direct probe of the neutral IGM, providing a unique three-dimensional view of the reionization process. Advanced radio interferometers such as the Low-Frequency Array \citep[LOFAR; e.g.,][]{mertens2020improved,Mertens2025lofar}, the Hydrogen Epoch of Reionization Array \citep[HERA; e.g.,][]{deboer2017hera,hera2022constraints,hera2023improved}, the Murchison Widefield Array \citep[MWA; e.g.,][]{tingay2013murchison,trott2020deep}, and the upcoming Square Kilometre Array \citep[SKA; e.g.,][]{koopmans2015cosmic} are designed to detect the faint 21-cm signal from the EoR. These observatories represent powerful tools for studying the growth and evolution of ionized bubbles, offering a window into the cosmic reionization process.

The low-frequency component of the SKA (SKA-Low) will be particularly powerful, providing direct images of the 21-cm signal \citep[e.g.,][]{mellema2015hi,giri2019tomographic}. Various techniques have been developed to extract critical information from these images, focusing on identifying neutral regions and characterizing their properties \citep[e.g.,][]{giri2018optimal,Gagnon-Hartman:2021erd,bianco2021deep,bianco2024deep,bianco2024serenet,gazagnes2021inferring}. \citet{giri2019neutral} demonstrated that neutral islands identified in 21-cm signal images can reveal key details about the reionization process, such as the topology of neutral island distribution. In the later stages of reionization, these neutral islands were relatively large (spanning more than 40 Mpc), 
making them easier to resolve in 21-cm signal images \citep{giri2019neutral,giri2024end}. In this study, we develop methods to map these large neutral islands, shedding light on the final phases of cosmic reionization.

In addition to the 21-cm signal, the \lya{} forest offers another powerful tool for probing the ionization state of the IGM \citep[see e.g.][for reviews]{becker2015review,fan2023quasars}. The \lya{} forest consists of a series of absorption features in the spectra of distant background sources, such as quasars and galaxies, caused by the presence of neutral hydrogen along the line-of-sight. As reionization progresses and the IGM becomes increasingly ionized, the opacity of the \lya{} forest decreases, allowing us to trace the evolving ionization state of the IGM. 
The residual \lya{} forest transmission in spectra probe regions within ionized bubbles, providing critical insights into the structure of the IGM during reionization \citep[e.g.,][]{gaikwad2020probing,Yang2020spikes,bosman2022hydrogen}. Traditionally, the \lya{} forest has been employed to study the post-reionized Universe, but recent advancements and deep spectroscopic campaign of high-redshift quasars \citep[e.g.][]{dodorico2023} have extended its applicability to higher redshifts, making it a valuable probe for the late stages of reionization \citep{becker2015evidence,becker2021mean,zhu2022darkgap,zhu2023mfp}.


Furthermore, \lya{} emission lines from galaxies embedded within reionized bubbles offer information about the ionization state of surrounding environment, with the size and extent of these bubbles playing a significant role in the observed strength, distribution and visibility of the emission \citep[e.g.,][]{Malhotra2006IGM,Dijkstra2011Lya,Mesinger2015Lya,mason2020measuring,Endsley2022strong,lu2024reionizing,lu2024mapping,nikolic2025mapping}. These \lya{} observations both in emission and absorption, when combined with other techniques such as 21-cm signal and spatially varying optical depth to the Cosmic Microwave Background (CMB), can provide a more comprehensive understanding of how ionized bubbles grew and merged during the final stages of the reionization epoch.

In this paper, we explore the capability of photometric IGM tomography to recover neutral islands and various cross-correlations as probes of reionization. Photometric IGM tomography utilizes deep narrow-band (NB) imaging to photometrically detect Ly$\alpha$ forest transmission along background galaxies \citep{mawatari2017imaging,kakiichi2023photometric}. This technique could potentially allow for the reconstruction of large-scale fluctuations in the IGM across a large area of the sky using existing wide-field imagers such as Subaru/Hyper-Suprime Cam (HSC). While the application of this photometric tomography technique to map highly ionized regions around quasars has previously been explored (\citealp{kakiichi2022photometric}, see also \citealt{schmidt2019lightecho}) along with a serendipitous photometric detection of transverse proximity effect of a quasar at $z\sim5.7$ \citep{bosman2020three}, the prospect of recovering neutral islands has not been exploited. The statistics of neutral islands and their sizes contain important information about reionization \citep{xu2017islandfast,giri2019neutral}. This would also open an alternative approach to examine the spatial correlation between galaxies and IGM opacity in 2D, complementary to previous studies along quasar sight-lines \citep{becker2018evidence,kashino2019evidence,ishimoto2022physical,Christenson2021,Christenson2023}. 

Photometric IGM tomography provides a complementary probe of the galaxy-IGM cross-correlation \citep{kakiichi2018role,meyer2020,garaldi2019crocs,garaldi2022thesan,Conaboy2025}. Recent JWST observations such as the EIGER \citep{kashino2023eiger} and ASPIRE \citep{wang2023aspire} surveys have validated the technique's ability to map the IGM structure during this epoch \citep{kashino2023eiger,kakiichi2025aspire}.  Crucially, the large-sky coverage afforded by photometric IGM tomography potentially makes it as an excellent tracer for cross-correlating with future 21-cm signal surveys, such as the SKA \citep{koopmans2015cosmic}, and Stage-IV CMB experiments \citep[e.g.,][]{abazajian2016cmb}. Such a detectable cross-correlation would be invaluable, serving as a robust validation tool and enabling a more comprehensive understanding of reionization.




This paper is structured as follows: Sec.~\ref{sec:modelling} outlines the methods used to model cosmic reionization and the corresponding observables. In Sec.~\ref{sec:igm_tomography}, we describe the photometric IGM tomography technique applied in this work. Sec.~\ref{sec:result} presents our results. We show the cross-correlation of the \lya{} forest transmissions with complementary datasets in Sec.~\ref{sec:cross_corr}. Finally, Sec.~\ref{sec:conclusion} concludes the paper.
Throughout this study, we assume a flat $\Lambda$-Cold Dark Matter cosmology with matter density contrast $\Omega_m=0.32$, baryon density contrast $\Omega_b=0.044$, normalized Hubble constant $h=0.673$, 
initial power spectrum index $n_s=0.96$ and matter clustering amplitude $\sigma_8 = 0.83$. This cosmology is consistent with the latest Planck CMB data \citep{planck2020cosmological}.

\section{Modelling Methods}\label{sec:modelling}
In this work, we consider large scale simulations of reionization to develop and test methods using IGM tomography to probe neutral islands and examine various cross-correlation signals during the end stages of reionization. In Sec.~\ref{sec:model_reion}, we present these simulations and, in Sec.~\ref{sec:model_lya_trans}, we describe the modeling of the \lya{} forest transmissions in these simulations.

\subsection{Simulating reionization}
\label{sec:model_reion}
We simulate the reionization of the IGM by post-processing dark matter only \nbody{} simulation snapshots with a fully numerical radiative transfer simulation framework, \pyccray{} \citep{hirling2023pyc}, that models the propagation of photon using a ray-tracing scheme \citep{mellema2006c2}. The \nbody{} simulation is performed using the \pkdgrav{} code \citep{potter2017pkdgrav3} in a simulation volume of length 298.5 Mpc in each direction, which is necessary to model the reionization of the IGM \citep[e.g.,][]{iliev2012can,giri2023suppressing}. This simulation consists of $2048^3$ particles, which are later assigned to grids of $256^3$ using a piece-wise spline mass assignment scheme \citep{sefusatti2016accurate}. The dark matter haloes are identified using the intrinsic friends-of-friends halo finder of \pkdgrav{}, and the smallest halo mass is about 10$^{9} M_\odot$. For more details about the simulation, we refer to \citet{giri2024end}.

\begin{figure*}
    \centering
    \includegraphics[width=0.99\linewidth]{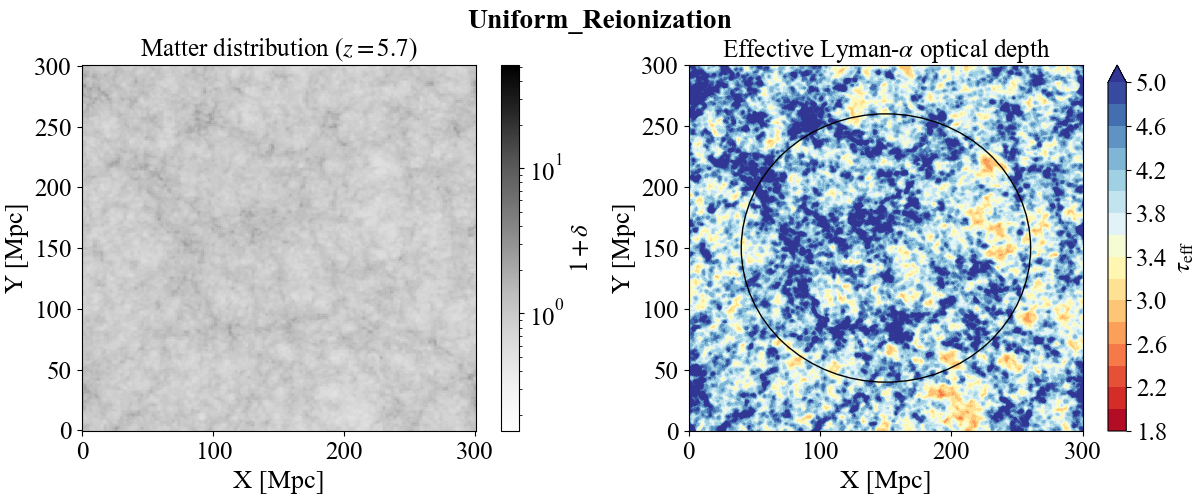}
    \includegraphics[width=0.99\linewidth]{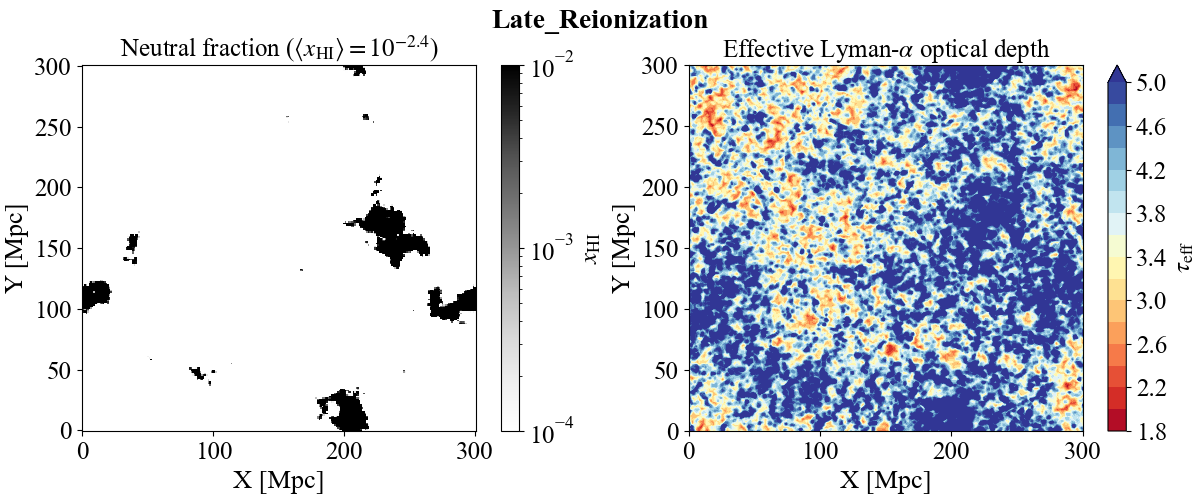}
    \vspace{-0.2cm}  
    \caption{
    Slices from the two reionization models at redshift $z=5.7$.  
    \textit{Top panels}: We show the matter density contrast slice on the left. In the \modelUni{} model, the universe has a uniform neutral fraction of $x_\mathrm{HI}=10^{-4}$ everywhere. On the right, we show the map of the effective optical depth of \lya{} photons $\tau_\mathrm{eff}$ at simulation resolution in the field of view direction. These values are averaged over 45 Mpc length scales along the line-of-sight. The black circle gives the field of view of the Subaru/HSC NB816 filter.
    \textit{Bottom panels}: We present a neutral fraction slice and the corresponding $\tau_\mathrm{eff}$ in the left and right, respectively, panels for the \modelSA{} model.
    In all the \lya{} effective optical depth maps, we can see \RefereeReport{that} the low values of transmissions correspond to the neutral hydrogen distribution.
    }
    \label{fig:models_tau_map}
\end{figure*}


In this study, we primarily focus on the end stages of reionization at $z=5.7$, where the corresponding HSC/NB816 filter is available to perform photometric IGM tomography.
We consider two reionization scenarios, which are the following:
\begin{itemize}
    \item \textbf{\modelUni{}}: We assume the universe to be uniformly ionized at $z\approx 6$ and, thus, has a neutral fraction $ x_\mathrm{HI}=10^{-4}$ everywhere at $z=5.7$. The top left panel of Fig.~\ref{fig:models_tau_map} shows the matter density contrast of a randomly chosen slice. This scenario represents a post-reionization universe where the ionization state of the IGM is sustained by the global ultraviolet (UV) background produced by ionizing photon sources. In this regime, the ionization state of the IGM becomes independent of the spatial distribution of ionizing sources, and the neutral hydrogen density follows the matter density field, exhibiting an \textit{outside-in} reionization pattern \citep[e.g.,][]{miralda2000reionization}.
    \item \textbf{\modelSA{}}: This reionization simulation, which was presented in \citet{giri2024end}, was calibrated to the latest high redshift measurements, such as the UV luminosity function. 
    \RefereeReport{To accurately model the highly ionized IGM relevant for the \lya{} forest, the non-equilibrium chemistry solver's convergence criterion was tightened, allowing the simulation to robustly track residual neutral fractions as low as $10^{-8}$. See \citet{hirling2023pyc} for the detailed algorithm. The ionizing background is generated by sources within the volume, and we use periodic boundary conditions for photon propagation. The impact of unresolved absorbers is included via a maximum travel distance for photons, following the redshift-dependent parametric form from \citet{worseck2014giant}, which was modified in \citet{giri2024end} to better agree with the latest observational data. This method successfully reproduces the observed evolution of the mean-free-path and the global UV background during the end stages of reionization \citep{giri2024end}.} The reionization extends to periods below $z \approx 6$, which is consistent with available observations of the \lya{} forest \citep[e.g.][]{choudhury2015lyman,kulkarni2019large,weinberger2019modelling,nasir2020observing,keating2020long,bosman2022hydrogen}. This simulation showed that there are very few but significantly large neutral islands (scales up to 40 Mpc) during the very late stages of reionization ($5 \lesssim z \lesssim 6$). The bottom left panel of Fig.~\ref{fig:models_tau_map} presents the neutral fraction map at an epoch with global neutral fractions $\langle x_\mathrm{HI} \rangle$ of $10^{-2.4}$ at $z=5.7$. 
    This panel shows several large neutral islands with dark color. One should note that the neutral fraction of these islands is not unity as the gas is partially ionized by the ionizing photons \citep{xu2017islandfast,giri2024end}.
\end{itemize}


\subsection{Lyman-$\alpha$ forest transmission}
\label{sec:model_lya_trans}
The transmission of the \lya{} flux through the IGM is modelled by employing the fluctuating Gunn-Peterson approximation \citep[FGPA; e.g.][]{Weinberg1997Hubble,rauch1998lyman} on sight-lines through our simulation volumes. The optical depth of these photons at every cell $i$ with neutral hydrogen number density $n_\mathrm{HI}$ per comoving volume is given by \citep[e.g.][]{choudhury2021studying,qin2021reionization},
\begin{eqnarray}
    \tau_i(z) = \kappa_\mathrm{res}\frac{\pi q_e^2}{m_e c} f_\alpha \lambda_\alpha \frac{(1+z)^3}{H(z)} n_\mathrm{HI,i} \ ,
    \label{eq:tau_i}
\end{eqnarray}
where $q_e$, $m_e$, and $c$ are the electron charge, electron mass and speed of light, respectively. $f_\alpha=0.416$ is the \lya{} oscillator strength and $\lambda_\alpha=1216$\AA~ is the \lya{} photon wavelength. $H(z)$ is the Hubble parameter value at redshift $z$. Finally, $\kappa_\mathrm{res}$ is a free parameter that models the impact of unresolved small scale neutral hydrogen distribution, which can be calibrated either with high-resolution simulations \citep[e.g.,][]{qin2021reionization} or observed data \citep[e.g.,][]{davies2016large}. 

For the \modelSA{} simulation used in this work, we estimate $\kappa_\mathrm{res}=0.56$ at $z=5.7$ by calibrating against the measurement from \citet{bosman2022hydrogen}. See Appendix~\ref{sec:kres_calibration} for more details on the calibration. We match the mean transmission from the simulation with the observed value. Additionally, we calibrate the $\kappa_\mathrm{res}$ for the \modelUni{} model to 0.16.
The latter case is less realistic as the cumulative probability distribution of $\tau_\mathrm{eff}$ has a steeper slope than \citet{bosman2022hydrogen} (comparison shown in Fig.~\ref{fig:kres_calibrated_S22}). However, we still perform a calibration to have a meaningful mean value and fluctuations in the \lya{} forest transmission map. 

Previous semi-numeric simulations applying FGPA required a subgrid model to estimate the residual neutral fraction in the ionized regions 
\citep[e.g.][]{davies2016large,qin2021reionization,choudhury2021studying}. These methods fully ionize the cells after ionization fronts have passed through them.
We do not apply such subgrid model as the state of the residual neutral fraction is calculated self-consistently by \pyccray{}, solving the radiative transfer equation in every cell.

The transmission of \lya{} flux is defined as $\mathcal{T} = \exp(-\tau_\mathrm{eff})$, where $\tau_\mathrm{eff}$ is the effective optical depth computed over the full width at half maximum (FWHM) of the NB816 filter. This corresponds to \RefereeReport{a} comoving length, 
\begin{eqnarray}
    L_{\rm NB}=\frac{c{\rm FWHM}}{H(z_{\rm NB})\lambda_{\alpha}}\approx42\rm\,Mpc,
\end{eqnarray}
where $\rm FWHM=113$\,\AA~ and $z_{\rm NB}=5.72$ for HSC/NB816 filter. 
In this study, we set this scale to 45 Mpc, 
which matches the NB filter width. Thus, the effective optical depth is given as
\begin{eqnarray}
    \tau_\mathrm{eff} = -\ln\left[\frac{1}{N}\sum_i\exp(-\tau_i)\right] \ ,
    \label{eq:tau_eff}
\end{eqnarray}
where the sum runs over the length of the NB filter, $N$ is the number of cells corresponding to the NB filter width, 
and $\tau_i$ is taken from Eq.~\ref{eq:tau_i}.
Throughout this paper, we assume the third axis of the reionization simulation to be along the line-of-sight. The evolution of the signal along the line-of-sight, known as light-cone effect, will have minor \RefereeReport{impact} within this $\sim$45 Mpc window \citep[e.g.][]{giri2018bubble} and, therefore, slices were taken from the simulated coeval cubes.

\subsection{Transmission maps} \label{sec:igm_tomo_maps}

We modelled the effective \lya{} optical depth ($\tau_\mathrm{eff}$) using Eq.~\ref{eq:tau_eff} on the reionization simulations presented in Sec.~\ref{sec:model_reion}. The right panels of Fig.~\ref{fig:models_tau_map} present the $\tau_\mathrm{eff}$, which is related to the \lya{} forest transmission $\mathcal{T}=\exp(-\tau_\mathrm{eff})$\footnote{In Fig.~\ref{fig:models_tau_map}, we do not plot the transmission directly, as the gradients in the map are difficult to visually interpret.}, for our two reionization models at $z=5.7$.
In the \modelUni{} model, which lacks neutral islands, the $\tau_\mathrm{eff}$ map does not feature any large, low-transmission structures. Instead, it closely follows the matter density fluctuations, which are shown in the top-left panel.
The bottom-right panel presents the \lya{} forest transmission map for the \modelSA{} model. Unlike the \modelUni{} model, this map does not follow the matter distribution but correlates with the neutral islands (bottom-left panel). For example, the large neutral island near \{X,Y\} $\simeq$ \{230,180\} Mpc is imprinted on transmission maps of \modelSA{} case as regions of high \lya{} optical depth. Therefore such transmission maps offer a valuable probe of reionization topology during its later stages.


\section{IGM tomography}\label{sec:igm_tomography}

In this section, we will give an overview of the mock observation of photometric IGM tomography. We first outline the methodology (Sec.~\ref{sec:igm_tomo_method}) and then discuss the observing strategy (Sec.~\ref{sec:obs_strategy}) explored for this study. Lastly, in Sec.~\ref{sec:struct_method}, we will present the pattern recognition technique employed to identify meaningful structure in noisy image data.

\subsection{Method} \label{sec:igm_tomo_method}



We map the IGM using photometric tomography technique described in \citet{kakiichi2022photometric,kakiichi2023photometric}. We refer the reader to these papers for a full description of the method. Briefly, photometric IGM tomography uses deep NB imaging to photometrically measure the residual \lya{} forest transmission along background galaxies. Intrinsic UV continua are estimated by the spectral energy density (SED) fitting to the broad-band photometry of the background galaxies. In this paper, we examine the observability of $z=5.7$ photometric IGM tomography using the Subaru/HSC NB816 filter. As the intrinsic UV continua can be determined from near-infrared data such as UltraVISTA \citep[][]{mccracken2012ultravista}, Euclid \citep[][]{Euclid2024overview}, and JWST NIRCam imaging, the dominant source of error in the observed IGM transmission is the photometric noise in the NB816 imaging. In the NB noise dominated regime, the observed \lya{} forest transmission along a $i$-th background galaxy is given by,
\begin{equation}
    \mathcal{T}_{{\rm obs},i} = \mathcal{T}_i+\delta \mathcal{T}_i,
\end{equation}
where $\mathcal{T}_i=\exp{(-\tau_{{\rm eff},i})}$ is the simulated \lya{} forest transmission computed from the effective optical depth (Eq.~\ref{eq:tau_eff}) over the NB filter width along a background galaxy sight-line at $\pmb{r}_i$ and $\delta\mathcal{T}_i$ is the observational noise, which is given by
\begin{equation}
    \delta\mathcal{T}=\frac{\delta f_{\rm NB}}{f_{{\rm UV},i}},
    \label{eq:obs_noise}
\end{equation}
where $\delta f_{\rm NB}$ is the 
noise of NB flux randomly drawn from the Gaussian distribution with the standard deviation corresponding to the $1\sigma$ limiting NB magnitude and $f_{{\rm UV},i}$ is the continuum flux of a background galaxy with an apparent UV magnitude $m_{{\rm UV},i}$.

We randomly sample the apparent UV magnitudes of background galaxies above the detection limit $m_{\rm UV}^{\rm limit}$ from the \cite{bouwens2021new} luminosity function. We generate $N$ background galaxies, $N=\Sigma_{\rm bkg}L_{\rm box}^2$, where $\Sigma_{\rm bkg}$ is the surface number density of background galaxies per comoving area,
\begin{equation}
    \Sigma_{\rm bkg}=\epsilon_{\rm spec}\int_{z_{\rm min}}^{z_{\rm max}}\frac{cdz}{H(z)}\int_{-\infty}^{M_{\rm UV}^{\rm limit}(m_{\rm UV}^{\rm limit},z)}\frac{dn}{dM_{\rm UV}}dM_{\rm UV} \ ,
    \label{eq:surface_num_dens}
\end{equation}
where $\epsilon_{\rm spec}$ is the spectroscopic confirmation rate of background galaxies and $M_\mathrm{UV}$ is the absolute magnitude.
For NB816 tomography, the maximum and minimum redshifts of background galaxies for which the \lya{} forest is covered by the NB816 filter is $z_{\rm min}=5.81$ and $z_{\rm max}=6.86$. Considering wide-field multi-object spectrographs such as Subaru/PFS and VLT/MOONS, the spectroscopic redshift of a larger number of background galaxies can be located by \lya{} emission lines. 
In this study, we assume a fiducial spectroscopic confirmation rate of $\epsilon_{\rm spec}=0.2$ (see e.g. \citealt{ouchiARA&Areview} and references therein), but will also consider the variation of this parameter. 

To reconstruct a 2D tomographic map of the IGM, we use a simple Gaussian kernel interpolation method as presented in \cite{kakiichi2023photometric}, 
\begin{equation}
    \mathcal{T}(\pmb{r})=\frac{\sum_{i=1}^N \mathcal{T}_i K_{l_s}(\pmb{r}-\pmb{r}_i)}{\sum_{i=1}^N K_{l_s}(\pmb{r}-\pmb{r}_i)},
    \label{eq:interp_Gaussian_kernel}
\end{equation}
where $K_{l_s}(\pmb{r}-\pmb{r}_i)$ is the 2D Gaussian kernel with the smoothing length (standard deviation) $l_s$. We set the smoothing length to be the typical separation between background galaxies, 
\begin{eqnarray}
    l_s = \frac{1}{\sqrt{\Sigma_\mathrm{bkg}}} \ ,
    \label{eq:l_smooth}
\end{eqnarray}
where $\Sigma_\mathrm{bkg}$ is given by Eq.~\ref{eq:surface_num_dens}. This quantity defines the resolution of the reconstructed \lya{} forest transmission map, which will be studied in Sec.~\ref{sec:result_num_bkg_src}.



\subsection{Observational requirement}\label{sec:obs_strategy}

\begin{figure}
    \centering
    \includegraphics[width=1.0\linewidth]{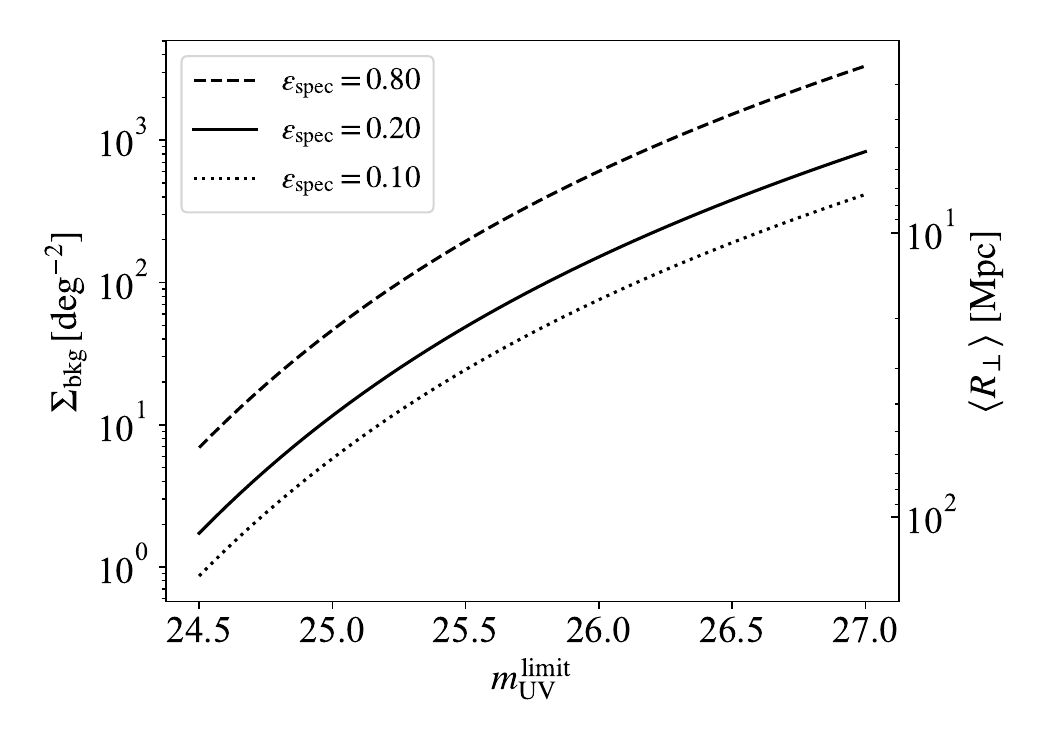}
    \vspace{-0.8cm}  
    \caption{
    The surface number density of background galaxies for photometric IGM tomography at $z=5.7$ as a function of the apparent limiting UV background of the background source population. The various curves indicate the spectroscopic confirmation rates of $\epsilon_{\rm spec}=0.8$ (dashed), $0.2$ (solid), and $0.1$ (dotted). The right y-axis shows the corresponding typical angular separation between background galaxies, which determines the spatial resolution of the tomographic map. 
    }
    \label{fig:bkg_galaxy_count}
\end{figure}

The observational requirements for IGM tomography is primarily set by the number of background galaxies, which determines the spatial resolution of the reconstructed IGM map. Fig.~\ref{fig:bkg_galaxy_count} shows the expected number density of background galaxies and the corresponding typical spatial resolution, $\langle R_\perp\rangle=\Sigma_{\rm bkg}^{-1/2}$. In order to recover neutral islands of sizes of $\sim10\rm \,Mpc$, we require background galaxies as faint as $m_{\rm UV}^{\rm limit}\approx25.5-26.0$ with a spectroscopic confirmation rate of $\epsilon_{\rm spec}=0.20$, providing the spatial resolution of $\sim10-20\rm ~Mpc$. A higher spectroscopic confirmation rate with brighter galaxies, such as those with $m_{\rm UV}^{\rm limit}=24.5$, similar to the Subaru/PFS survey \citep{Greene2022}, which will provide a spatial resolution of $\sim50\rm ~Mpc$. While this is sufficient to recover the very large neutral islands extending over 100 Mpc \citep{becker2015evidence,keating2020long}, higher spatial tomographic resolution is desirable to map smaller, more common neutral islands at the end of reionization. 

Once the spatial resolution, i.e., the required apparent UV magnitude, is chosen, the requirements for NB depth and other broad-band filters follow naturally. A rule-of-thumb is that NB depth needs to be, assuming a flat UV continuum for background galaxies,
\begin{equation}
m_{\rm NB}^{\rm limit} \approx m_{\rm UV}^{\rm limit} - 2.5 \log_{10} e^{-\tau_{\rm eff}(z)}
\end{equation}
to photometrically detect the \lya{} forest transmission along individual background galaxies with $m_{\rm UV} < m_{\rm UV}^{\rm limit}$. The effective optical depth at mean IGM is $\tau_{\rm eff} = 3.5$ at $z = 5.7$ \citep{bosman2022hydrogen}. The effective optical depth measured over the FWHM of the NB816 filter spans between $\tau_{\rm eff} =2.4$ and $<7.3$ along quasar sight-lines \citep{Christenson2023}. Our \texttt{Late\_Reionization} simulation spans a similar range of $\tau_\mathrm{eff}$ (see Appendix \ref{sec:kres_calibration}). This means that for a background galaxy with $m_{\rm UV} = 25.0$ ($26.0$), an NB depth of $m_{\rm NB} \simeq 27.5$ ($28.5$) is required for the detection of transmissive IGM sight-lines. This seems to set quite a stringent requirement for the NB depth to map the entire IGM, including neutral islands that will have very high effective optical depth. However, as we show in Sec.~\ref{sec:result}, thanks to numerous background galaxies that help reduce the noise level, we can achieve a sensible reconstruction of the IGM tomographic map and recover the properties of neutral islands using extremely deep NB imaging with a depth of $27.5 \, (3\sigma)$.

\subsection{Recovering the neutral islands from IGM tomography}\label{sec:struct_method}

\begin{figure*}
    \centering
    \includegraphics[width=1.0\linewidth]{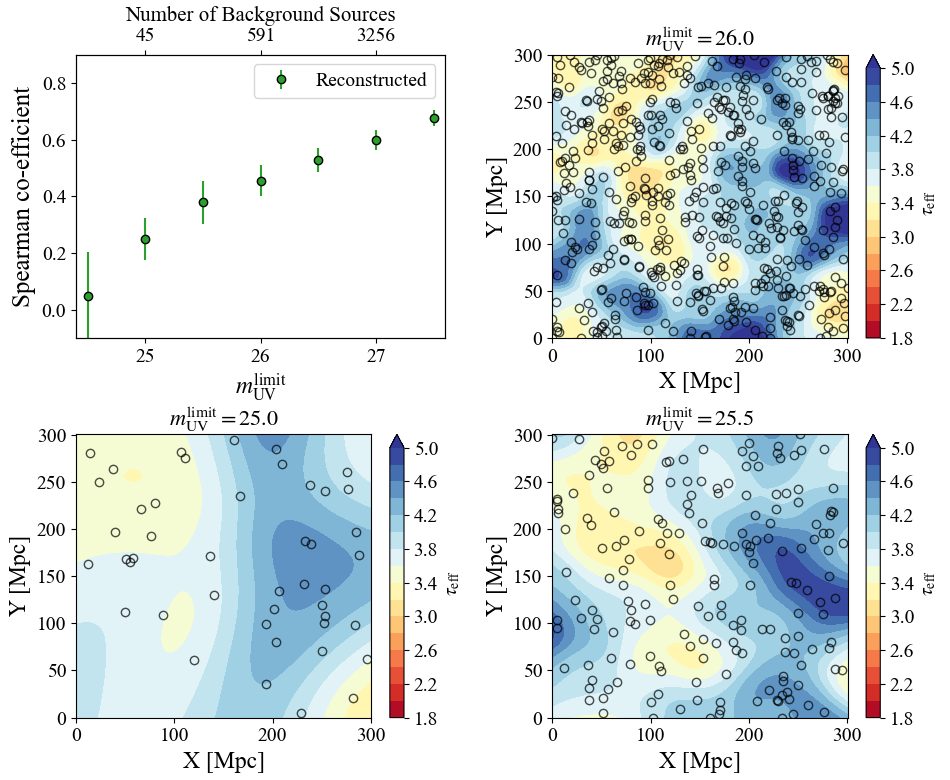}
    \vspace{-0.4cm}  
    \caption{
    The number of background galaxies required for IGM tomography. \textit{Top-left panel}: The Spearman correlation coefficient of the ground truth with the reconstructed transmission maps applying different brightness ($m_\mathrm{UV}$) cuts on the luminosity function. The x-axis on the top gives the number of background sources corresponding to these cuts. We see that the correlation improves with an increase in the number of background sources.
    \textit{Top-right to bottom-left in clockwise}: The reconstructed \lya{} effective optical depth maps employing brightness limits at $m_\mathrm{UV}=26$, 25.5 and 25, respectively. The black circles represent the positions of the background sources. The resolution of the maps decreases with the number of background sources, which causes a decrease in correlation with the ground truth.
    }
    \label{fig:Source1_SinkA_number_bkg_sources}
\end{figure*}

The bottom panels of Fig.~\ref{fig:models_tau_map} show that the last remaining neutral islands leave distinct imprints on the effective \lya{} optical depth maps, appearing as regions of low \lya{} photon transmission. We can therefore identify these neutral islands by detecting such patterns in the reconstructed \lya{} forest transmission maps. However, since the reconstructed images are expected to be noisy (see Sec.\ref{sec:feasibility}), we require a robust pattern recognition technique that works effectively on images with low signal-to-noise ratio (SNR).

In this study, we employ a modified pattern recognition approach that combines superpixel techniques with a threshold-based method. Superpixel algorithms are designed to over-segment images and are commonly used to identify the boundaries of physical structures \citep[see e.g.,][for a review]{ibrahim2020image}. We use the Simple Linear Iterative Clustering (SLIC) algorithm \citep{achanta2012slic} to identify superpixels in our maps—groups of pixels with similar properties. This algorithm has previously been adapted to analyze noisy 21-cm signal images \citep{giri2018optimal} and used to study the evolution of ionized and neutral regions \citep{giri2018bubble,giri2021measuring}. For IGM transmission maps, SLIC relies on the magnitude of $\mathcal{T}$ and the spatial connectivity of pixels.

While the superpixel map captures the boundaries of physical structures, the structures themselves remain fragmented into smaller pixel clusters. To recover the full, continuous structure, we need a property that allows us to merge these superpixels. In \citet{giri2018optimal}, distinct histogram features were used to identify ionized regions. However, due to the high level of photometric noise in reconstructed \lya{} transmission maps (see Sec.~\ref{sec:result_num_bkg_src}), histogram features become indistinguishable and are thus not usable in our case.
Instead, we identify a threshold value on the superpixel-segmented image that minimizes the cross-entropy between the foreground (neutral islands) and background. This thresholding technique was proposed by \citet{li1998iterative}. A detailed description of our modified structure-finding algorithm, along with a toy example, is provided in Appendix~\ref{sec:struct_ident}.

\section{Result}\label{sec:result}

We now present results forecasting the capabilities of photometric IGM tomography. First, in Sec.~\ref{sec:result_num_bkg_src}, we examine the number of background sources required to map the IGM. Then, in Sec.~\ref{sec:feasibility}, we assess the feasibility of using reconstructed maps to retrieve physical structures in the presence of noise. Finally, in Sec.~\ref{sec:result_isd}, we evaluate the ability to detect neutral islands.

\subsection{Reconstructed IGM tomographic maps}\label{sec:result_num_bkg_src}

The reconstruction of the \lya{} forest transmission maps is highly dependent on the instrument's ability to observe background sources. 
To evaluate the required number of background sources, 
we employed the Spearman correlation coefficient \citep{spearman1961proof}, which measures the degree of correlation between the reconstructed maps and the true underlying signal. The Spearman coefficient ranges from -1 (perfect anti-correlation) to 1 (perfect correlation). This analysis was performed using the \modelSA{} reionization model, correlating the reconstructed maps with the true transmission maps (bottom-right panel of Fig.~\ref{fig:models_tau_map}).

In the top-left panel of Fig.~\ref{fig:Source1_SinkA_number_bkg_sources}, we show how the Spearman coefficient varies with varying $m^\mathrm{limit}_\mathrm{UV}$ defining the background source population. 
The top x-axis of the plot indicates the corresponding number of background sources for each magnitude limit marked on the bottom x-axis. This number is derived from the surface number density of the background galaxies, which is defined by the magnitude limit $m^\mathrm{limit}_\mathrm{UV}$ (see Eq.~\ref{eq:surface_num_dens}). The error bars shown in this plot were derived by performing the analysis on 25 different randomly selected slices from our simulation. The results show that the quality of the reconstruction improves as the number of background sources increases. For the remainder of this study, we adopt a magnitude cut of $m^\mathrm{limit}_\mathrm{UV} = 26$, which corresponds to a Spearman coefficient of $\sim$0.4.

For a clearer understanding of the performance of our reconstruction,
Fig.~\ref{fig:Source1_SinkA_number_bkg_sources} includes reconstructed \lya{} transmission maps for $m^\mathrm{limit}_\mathrm{UV} = 26$ (top-right), $25.5$ (bottom-right), and $25$ (bottom-left). These maps also display the positions of background sources with black circles. As \RefereeReport{is} evident, the resolution of the reconstructed maps decreases as the number of background sources is reduced, which is defined by the smoothing scale given in Eq.~\ref{eq:l_smooth}. This results in the smearing of small-scale features present in the ground truth, which explains the corresponding decline in the Spearman coefficient.

\begin{figure}
    \centering
    \includegraphics[width=1.0\linewidth]{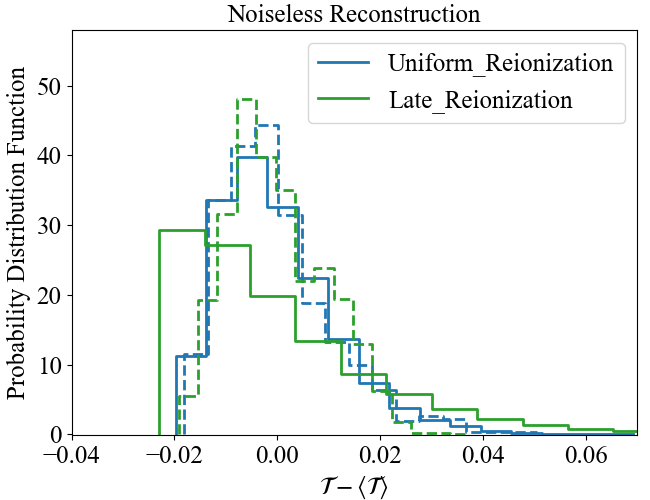}
    \vspace{-0.4cm}  
    \caption{
    The normalized histogram of the \lya{} forest transmission $\mathcal{T}$, which is shown in the right panels of Fig.~\ref{fig:models_tau_map}. The mean value of the map $\langle \mathcal{T}\rangle$ is subtracted before estimating these histogram to focus on their shape. The solid curves represent the ground truth of the two reionization models, while the dashed lines show the histogram of the reconstructed transmission maps assuming background galaxies down to $m_{\rm UV}^{\rm limit}=26.0$. The peak of the histogram for the simulation with neutral islands (\modelSA{}) shifts towards lower transmission values. The reconstructed \lya{} transmission maps exhibit more skewness for the \modelSA{} compared to \modelUni{} model. 
    }
    \label{fig:models_tau_hist}
\end{figure}

The imprint of neutral islands in the tomographic map can also be observed in the one-point statistics. Fig.~\ref{fig:models_tau_hist} shows the distribution of pixel values, $\mathcal{T} - \langle T \rangle$, in the \lya{} forest transmission maps. 
The solid curves correspond to the true \lya{} forest transmission maps (right panels of Fig.~\ref{fig:models_tau_map}). All the histograms are asymmetric, highlighting the non-Gaussian nature of the fields. The presence of neutral islands in the \modelSA{} model produces large regions of low \lya{} transmission in the IGM transmission map, leading to low pixel values and high skewness in the one-point statistics. To verify the robustness of this feature, we calculate the skewness across 25 randomly selected slices from the simulations, finding $1.71 \pm 0.12$ for the \modelSA{} model, compared to $1.01 \pm 0.11$ for the \modelUni{} model.

Unfortunately, the reconstruction process makes it challenging to clearly distinguish between the models using one-point statistics, as illustrated by the dashed lines in Fig.~\ref{fig:models_tau_hist}. Although the general trend persists in the reconstructed IGM maps—even in the presence of noise—the histograms appear less skewed compared to the ground truth. The difference between the models is also significantly reduced, with skewness values of $0.66 \pm 0.35$ for the \modelSA{} model and $0.52 \pm 0.33$ for the \modelUni{} model. This reduction in skewness is primarily due to the smoothing introduced by the Gaussian kernel interpolation used in the reconstruction (see Eq.~\ref{eq:interp_Gaussian_kernel}). \RefereeReport{As seen in Fig.~\ref{fig:models_tau_hist}, the impact of this interpolation is more pronounced on the \modelSA{} model, as its highly skewed true PDF—arising from presence of neutral islands—is more susceptible to the smoothing effect.} 

\RefereeReport{This analysis shows that it is difficult to robustly extract the imprint of neutral islands using one-point statistics alone, and this is before considering the impact of observational noise, which would only further diminish the distinguishing power of this metric.} This limitation of one-point statistics highlights the need for complementary analyses that leverage spatial information—either through the identification of morphological features such as neutral islands, or through spatial cross-correlations with external tracers such as galaxies, 21-cm signal maps or the optical depth to the CMB. In the following sections, we show that such spatial analyses (Sec. \ref{sec:result_isd}) and cross-correlation techniques (Sec. \ref{sec:result_cross_galaxies}–\ref{sec:result_cross_CMB}) can indeed recover valuable information from photometric IGM tomographic maps.

\subsection{Noisy IGM tomographic maps: feasibility}\label{sec:feasibility}



\begin{figure*}
    \centering
    \includegraphics[width=0.99\linewidth]{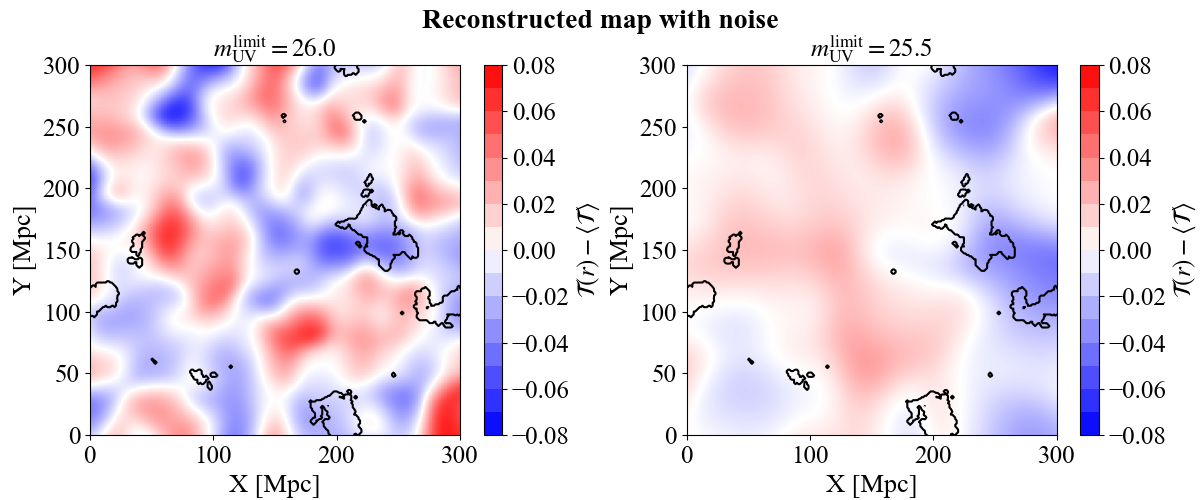}
    \includegraphics[width=0.99\linewidth]{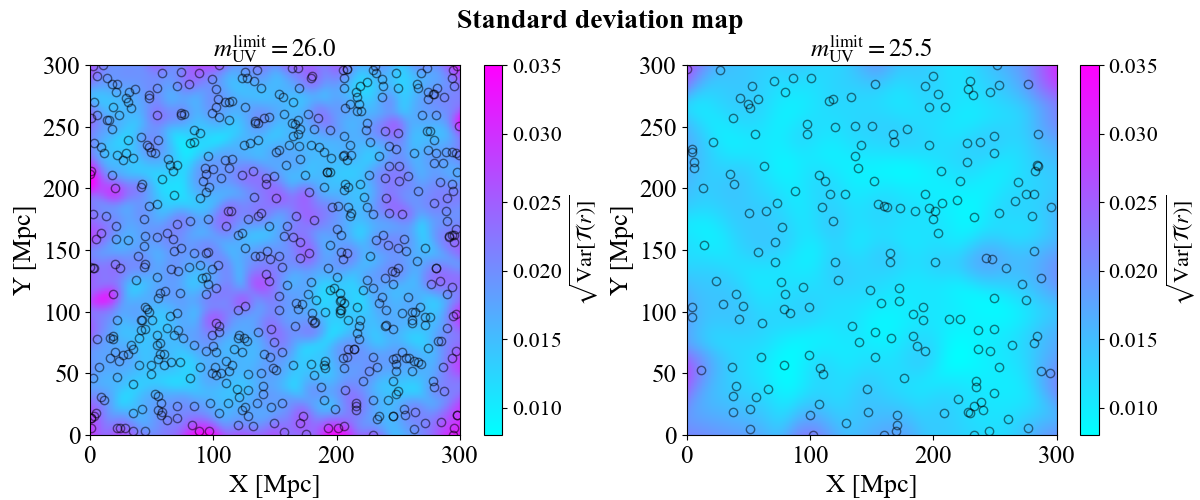}
    \includegraphics[width=0.99\linewidth]{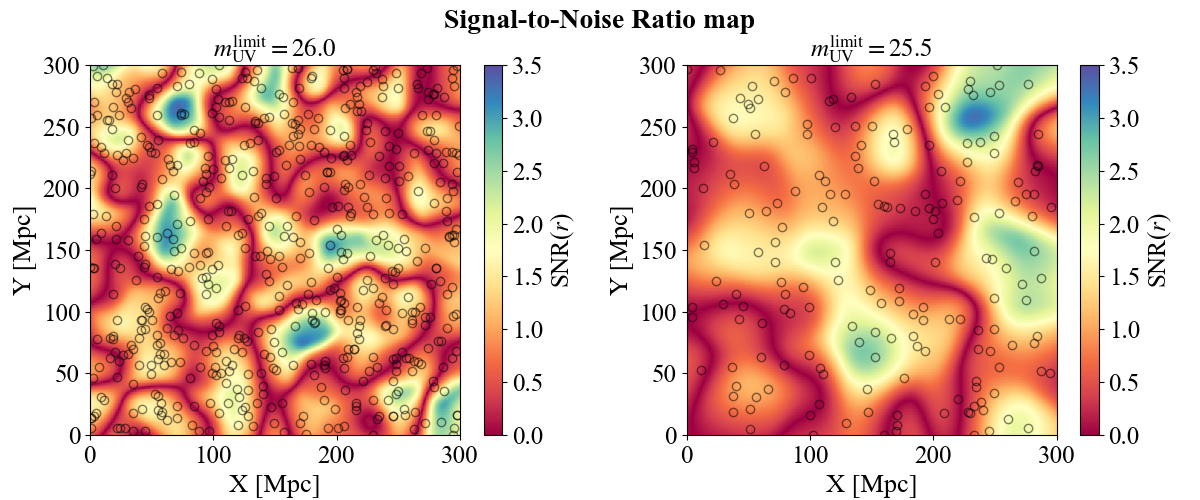}
    \vspace{-0.2cm}  
    \caption{
     Feasibility of recovering coherent IGM features from photometric IGM tomographic maps. We assume a $3\sigma$ narrow-band (NB) depth of $n_{\rm NB}^{\rm limit} = 27.5$. The limiting UV magnitudes are $m_{\rm UV}^{\rm limit} = 26.0$ (left panels) and $25.5$ (right panels). The top, middle, and bottom rows show the noisy reconstructed IGM tomographic maps, the corresponding standard deviation maps, and the signal-to-noise ratio (SNR) maps, respectively. The \RefereeReport{contours of the neutral islands at simulation resolution is shown in the top panel and} simulated positions of background galaxy sight-lines are indicated by black circles \RefereeReport{in the bottom two panels.}
    }
    \label{fig:feasibility}
\end{figure*}

\begin{figure*}
    \centering
    \includegraphics[width=0.49\linewidth]{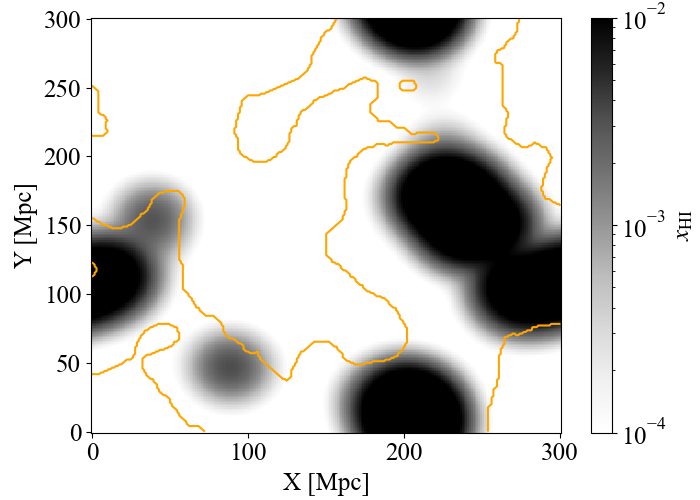}
    \includegraphics[width=0.49\linewidth]{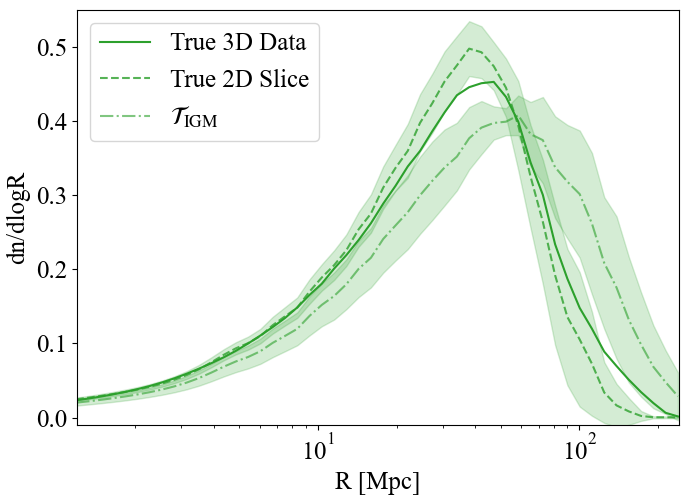}
    \vspace{-0.2cm}  
    \caption{
    Neutral islands probed by the \lya{} forest transmission maps in our \modelSA{} model. 
    \textit{Left panel}: The neutral fraction map, identical to the one in the bottom-left panel of Fig.~\ref{fig:models_tau_map}, but smoothed to match the resolution of the reconstructed maps. Orange contours indicate the boundaries of neutral regions identified in the corresponding reconstructed \lya{} forest transmission map using background galaxies down to $m_{\rm UV}^{\rm limit}=26.0$ and the NB depth of 27.5 ($3\sigma$). These contours closely trace the large neutral islands with $x_\mathrm{HI}\gtrsim 10^{-3}$.
    \textit{Right panel}: The size distribution of neutral islands in the reionization simulation. Solid and dashed lines represent the 3D and 2D neutral fraction data, respectively, both smoothed to match the \lya{} forest transmission map resolution. Dash-dotted lines show the size distribution of neutral islands identified in the transmission maps, demonstrating that these maps effectively capture the typical island sizes.
    }
    \label{fig:Source1_SinkA_bubble_map}
\end{figure*}

To assess the feasibility of identifying the ionized regions and neutral islands from noisy IGM tomographic maps, we examine the noise properties of the reconstructed maps. The variance map can be estimated by
\begin{equation}
   {\rm Var}\left[\mathcal{T}(\pmb{r})\right]=\frac{\sum_{i=1}^N  K_{l_s}(\pmb{r}-\pmb{r}_i)^2 \delta\mathcal{T}^2_i}{\left[\sum_{i=1}^N K_{l_s}(\pmb{r}-\pmb{r}_i)\right]^2},
\end{equation}
where $\delta\mathcal{T}_i$ is the observation\RefereeReport{al} noise on the measured IGM transmission along a background galaxy with $m_{{\rm UV},i}$ and for the NB limiting magnitude $m_{\rm NB}^{\rm limit}(1\sigma)$. Fig.~\ref{fig:feasibility} shows the reconstructed noisy IGM tomographic map (top panels) along with the maps of variance (middle panels) and SNR (bottom panels). We subtracted the mean value $\langle \mathcal{T}\rangle$ from \RefereeReport{the} reconstructed noisy IGM tomographic map. In the left panels, we assume the limiting UV magnitude of $m_{\rm UV}^{\rm limit}=26.0$ and a $3\sigma$ NB depth of $m_{\rm NB}^{\rm limit}=27.5$. We estimate the SNR using the noisy reconstructed IGM tomographic map to mimic real observations, which is defined as
\begin{equation}
    {\rm SNR}(\pmb{r})=\frac{\left|\mathcal{T}(\pmb{r})-\langle\mathcal{T}\rangle\right|}{\sqrt{{\rm Var}\left[\mathcal{T}(\pmb{r})\right]}},
\end{equation}
where $\langle\mathcal{T}\rangle$ is the mean IGM transmission measured using all background galaxies. 

\RefereeReport{The top panels of Fig. \ref{fig:feasibility} compare the results from the deeper survey (left) with the shallower one (right). To facilitate a direct visual assessment of the reconstruction quality, we overlay black contours that trace the true boundaries of the neutral islands from our simulation. The higher fidelity of the deeper survey is evident; for example, the left panel shows the successful recovery of the low-transmission region corresponding to the neutral island near \{X,Y\} = \{200,20\} Mpc. In contrast, the shallower map on the right illustrates the impact of noise, as a spurious high-transmission signal appears within this same neutral island. It is important to remember that these maps represent just a single realization of observational noise. For this reason, the variance and SNR maps (middle and bottom panels) provide a more statistically robust assessment of the reconstruction's capabilities across the field.}

\RefereeReport{The middle panels of} Fig. \ref{fig:feasibility} shows that the variance in the reconstructed IGM transmission map is lower in regions with bright and/or numerous background galaxies. This is expected, as areas with more sight-lines to background galaxies allow the noisy measurements to be averaged out more effectively. While the overall SNR remains modest (typically $\rm SNR \sim 3$), some excess transmissive regions of the IGM can still be identified as SNR peaks (for e.g., near \{X,Y\} = \{60,160\} Mpc and \{170,80\} Mpc). These areas correspond to highly ionized regions in the true IGM transmission map (see the bottom panels of Fig.~\ref{fig:models_tau_map}). Conversely, the uncertainty increases in areas with sparse background sources and near the edges, where there are no background sight-lines beyond the boundary.

Interestingly, some opaque regions of the IGM can also appear as peaks in the SNR map. For example, the region at \{X,Y\} $\simeq$ \{200,150\} Mpc, associated with neutral islands, shows up as a modest SNR peak. This is because the significance of transmission measurements along individual sight-lines depends on the depth of the NB observation relative to the UV magnitude, approximately given by $\tau_{\rm eff} \approx m_{\rm NB}(1\sigma) - m_{\rm UV}$. Hence, regions with a high density of bright background galaxies also aid in identifying opaque regions in the IGM.

However, the identification of both ionized bubbles and neutral islands is affected by observational noise. As shown in Fig.~\ref{fig:feasibility}, some high SNR peaks in the reconstructed maps are spurious and are not associated with the true ionized bubbles or neutral islands. We can mitigate this at the cost of spatial resolution by limiting the background sources to only the brighter ones. The right panels of Fig.\ref{fig:feasibility} show the reconstructed map and corresponding SNR after limiting the background sources to $m_{\rm UV} < 25.5$. In this case, we can clearly see a better identification of the coherent IGM features with less contamination from the spurious noise peaks. Thus, photometric IGM tomography facilitates the identification of bubble and neutral island candidates through spatial morphological analysis.

\subsection{Identifying neutral islands} \label{sec:result_isd}

To identify neutral islands, we apply the structure identification algorithm described in Sec.~\ref{sec:struct_method} to the noisy reconstructed IGM transmission map. As shown in the left panel of Fig.\ref{fig:Source1_SinkA_bubble_map}, the algorithm successfully recovers the large-scale features of the neutral islands. The detected regions, outlined by orange contours, trace the underlying neutral fraction map from the \modelSA{} simulation, smoothed to match the resolution of the reconstructed \lya{} forest transmission maps. Our results demonstrate that large-scale neutral islands can be identified, despite some imperfections caused by noise in the reconstructed maps. While increasing the number of background sources could improve spatial resolution, it would also introduce more noise—highlighting a trade-off between resolution and reliability.


To further explore the statistical properties of the neutral islands, we \RefereeReport{apply} the mean free path algorithm\footnote{This algorithm traces rays in random directions from points within the neutral islands, stopping at the first boundary encountered. We used approximately $10^7$ random points to create a probability distribution of the ray lengths. For more details, see \citet{giri2018bubble} and \citet{giri2019neutral}.} introduced by \citet{mesinger2007efficient} to estimate size distribution. In the right panel of Fig.~\ref{fig:Source1_SinkA_bubble_map}, we compare the island size distribution measured from the reconstructed \lya{} forest transmission maps (dash-dotted line) with simulations: the three-dimensional (3D) neutral fraction coeval cube (solid line) and a two-dimensional (2D) slice from the ionization map smoothed over 45 Mpc along the line-of-sight (dashed line). The uncertainty is estimated from the scatter of the island size distributions from 25 random slices, reflecting the sample variance. All data were smoothed with a Gaussian filter of size $l_\mathrm{s}$ for consistency, as smoothing is crucial when comparing size distributions \citep{giri2018bubble}. For the true data, we assumed a threshold of $x_\mathrm{HI} = 10^{-3}$ on the simulated neutral fraction map.


Fig.~\ref{fig:Source1_SinkA_bubble_map} shows that the measured island size distribution from the noisy IGM tomographic maps follows the true 3D distribution with an accuracy of approximately 0.3 dex. \RefereeReport{The recovered neutral island size distribution peaks around 50 Mpc, which is in broad agreement with the true underlying value. However, there are notable quantitative differences between the recovered 2D distribution and the true 3D one. The recovered distribution is significantly broader, and its peak is shifted towards larger scales compared to the 3D case. This discrepancy is an expected consequence of two effects. First, projecting a complex 3D field onto a 2D slice inherently alters the size distribution of the field. Second, the noisy reconstruction process tends to blur the boundaries of islands, sometimes merging distinct, nearby regions into single, larger apparent structures, which likely explains the observed bias. Therefore, while the recovered 2D distribution should not be interpreted as a direct, one-to-one proxy for the 3D island sizes, our analysis serves as a crucial proof-of-concept. It demonstrates that the characteristic scale of neutral islands is accessible and that their size distribution is a measurable quantity with photometric tomography.} A more detailed analysis connecting statistical information from 3D and 2D data, using stereological methods \citep[e.g.,][]{elias1971stereology}, will be addressed in future work.




\section{Cross-correlation studies}\label{sec:cross_corr}

In this section, we explore the cross-correlations of the \lya{} forest transmissions and other datasets.
First, we study the cross-correlation of the \lya{} forest transmission with galaxy distribution in both real (Sec.~\ref{sec:result_mean_transmission_galaxies}) and Fourier (Sec.~\ref{sec:result_cross_galaxies}) space.
Next, we present the cross-correlation of the \lya{} forest transmission maps with the 21-cm signal in Sec.~\ref{sec:result_cross_21cm}.
Finally, in Sec.~\ref{sec:result_cross_CMB}, we explore the correlation of \lya{} forest transmissions with the patchy optical depth which can be obtained from CMB measurements.



\subsection{Mean \lya{} forest transmission around galaxies}\label{sec:result_mean_transmission_galaxies}

\begin{figure}
    \centering
    \includegraphics[width=1.0\linewidth]{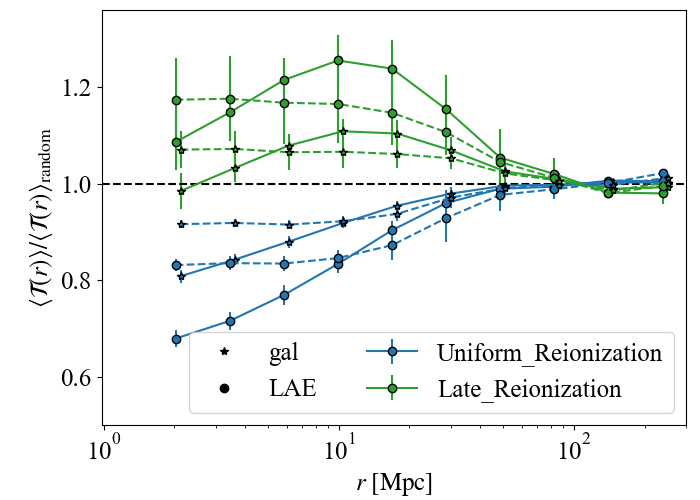}
    \vspace{-0.4cm}  
    \caption{
    The average transmission of the \lya{} forest around galaxies is presented for two reionization models. The curves are normalized by the transmission around randomly distributed galaxies, representing a case with no correlation. Filled stars and circles represent two galaxy survey catalogs: the full galaxy catalog (gal) and the subset containing massive galaxies (LAE). The solid and dashed lines \RefereeReport{are} estimated from simulated and reconstructed transmission maps, respectively. For clarity, the average transmission for the `gal' dataset is shifted slightly to the right by 5 per cent on the logarithmic scale. The \modelSA{} model shows higher transmission around galaxies compared to randomly distributed galaxies, while the \modelUni{} model exhibits lower transmissions. At large distances, the value converges to the mean background transmission. 
    }
    \label{fig:average_profile_around_galaxies}
\end{figure}

We study the average transmission of the \lya{} forest around the galaxies, which can be quantified as \citep{kakiichi2023photometric},
\begin{eqnarray}
    \langle\mathcal{T}(r)\rangle = \frac{1}{N_\mathrm{pairs}(r)} \sum^{N_\mathrm{g}}_{j=1} \sum^{N_\mathrm{bg}}_{i=1} \mathcal{T}_i ~\delta_\mathrm{D}(\pmb{r}-\pmb{r}_{ij}) \ ,
\end{eqnarray}
where $N_\mathrm{pairs}$ is the number of pairs in each angular bin. $\delta_\mathrm{D}$ is the Dirac's function that gives a value of unity when the angular distance $\pmb{r} \approx \pmb{r}_{ij}$ where $\pmb{r}_{ij}$ is the distance of the pixel from the $i^\mathrm{th}$ galaxy. $N_\mathrm{g}$ and $N_\mathrm{bg}$ are the number of galaxies in the observed field and the background galaxies producing the \lya{} spectra, respectively. 
We refer interested readers to appendix A of \citet{kakiichi2023photometric} for a full derivation of this estimator. The variance on the average \lya{} forest transmission is given as,
\begin{eqnarray}
    \sigma^2_\mathcal{T}(r) = \frac{1}{N^2_\mathrm{pairs}(r)} \sum^{N_\mathrm{g}}_{j=1} \sum^{N_\mathrm{bg}}_{i=1} \sigma^2_{\mathcal{T}_i} ~\delta_\mathrm{D}(\pmb{r}-\pmb{r}_{ij}) \ ,
\end{eqnarray}
where $\sigma^2_{\mathcal{T}}$ is the variance of the noise in the reconstructed map (see Eq.~\ref{eq:obs_noise}). The cumulative variance is $\sigma^2_\mathcal{T}+\sigma^2_\mathrm{CV}$, where $\sigma^2_\mathrm{CV}$ is cosmic variance determined by reconstructing the transmission map of 25 randomly selected slices from our simulations.

Here, we investigate the cross-correlation between \lya{} forest transmission maps and the spatial distribution of high-redshift galaxies in the NB redshift slice, which is $z\sim 5.7$. We simulated the following galaxy datasets:
\begin{itemize} 
    \item Random: We assume galaxies are distributed randomly in the simulation volume and therefore are uncorrelated with the matter distribution. This dataset represents the reference case where the \lya{} forest transmission profiles are uncorrelated with the realistic galaxy distribution.
    \item All galaxies (gal): This dataset includes the complete catalog of galaxies available in our simulation, containing photon sources hosted by dark matter halos down to a mass of $\sim 10^9\rm\,M_\odot$. 
    \item \lya{} emitters (LAE): This dataset comprises only luminous \lya{} emitting galaxies, hosted by dark matter halos with masses $\gtrsim 10^{11}\rm\,M_\odot$, mimicking a survey dataset of \lya{} emitter \citep[e.g.,][]{ouchi2010statistics,ouchi2018systematic}. These \lya{} emitters are typically found in massive halos, as these galaxies need to reside within sufficiently large ionized regions to exhibit bright \lya{} photon emission \citep[e.g.,][]{dayal2012lyalpha,hutter2023astraeus}. 
\end{itemize}
It is important to note that these galaxy samples differ from the background galaxies or sources used to construct the \lya{} forest transmission maps.

In Fig.~\ref{fig:average_profile_around_galaxies}, we present the average \lya{} forest transmissions around galaxies in the two datasets considered here, normalized by the case where galaxies are randomly distributed. This quantity measures the cross-correlation between galaxies and the \lya{} forest transmission \citep{kakiichi2023photometric}. The solid curves indicate the galaxy-\lya{} forest cross-correlation measured from direct \lya{} forest transmission measurements $\mathcal{T}_i$ from background galaxies, whereas the dashed curves indicate those measured using the reconstructed \lya{} forest transmission map. 

We find that the \modelSA{} model shows excess \lya{} forest transmission \RefereeReport{in the vicinity of galaxies compared to the random case. This behavior is expected, as galaxies typically reside in highly ionized regions, and it highlights the positive correlation between the transmission maps and the galaxy distribution in the presence of neutral islands.} 
At small radii ($r\lesssim 1$ Mpc), the average \lya{} forest transmission decreases due to the local matter overdensity around galaxies. However, the small-scale drop in the transmission profile is not observed in the cross-correlation measured from the reconstructed map (green dashed lines) because of the effect of smoothing during the reconstruction process. The overall shape of the cross-correlation is consistent with what has been found in previous studies \citep[e.g.,][]{garaldi2022thesan,Conaboy2025}. The galaxy-\lya{} forest cross-correlation from photometric IGM tomography shows an overall excess because the line-of-sight smoothing from the NB filter smooths out fluctuations smaller than the NB filter width of $ \sim 40 \, \rm Mpc$. The excess \lya{} forest transmission estimated with the full galaxy catalogue is lower than the LAE case because low-mass galaxies can be found in less ionized regions, such as voids, where neutral islands exist. 


Additionally, Fig.~\ref{fig:average_profile_around_galaxies} shows that the average \lya{} forest transmission around galaxies in the \modelUni{} model is lower than in the corresponding random case.
This behavior arises from the \textit{outside-in} reionization topology/density fluctuations in these models, where dense regions remain more neutral around galaxies. 
Consequently, we observe an excess \lya{} forest absorption as we go small radii. 
The curves corresponding to the reconstructed \lya{} forest transmission maps (blue dashed lines) show a flattening at small radii due to the effect of smoothing during the reconstruction process. 
The average transmission around galaxies in the LAE catalogue shows more excess absorption than one with the full galaxy catalogue. This is because massive galaxies are more biased and preferentially trace dense regions, which remain more neutral under the \textit{outside-in} reionization topology. Thus, the lower average transmission compared to the random case is a robust signature of the end of reionization, marked by the completion of reionization and the absence of any neutral islands in the IGM.

\subsection{Cross-power spectrum with galaxy distribution}\label{sec:result_cross_galaxies}

\begin{figure}
    \centering
    \includegraphics[width=1.0\linewidth]{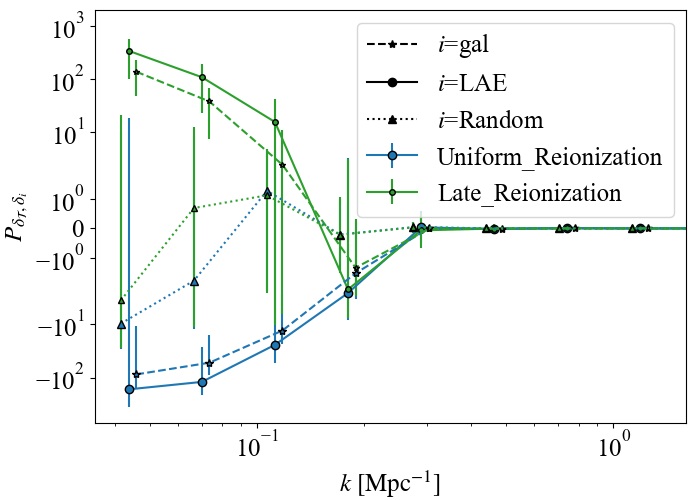}
    \vspace{-0.4cm}  
    \caption{
    Cross-power spectra of the reconstructed \lya{} forest transmission maps and galaxy distribution are presented for two reionization models. Filled stars and circles represent two galaxy survey catalogs: the full galaxy catalog (gal) and the subset of massive galaxies (LAE). The dotted lines with triangle symbols correspond to a scenario where galaxies are randomly distributed. Error bars indicate the impact of sample variance and reconstruction noise. For clarity, the cross-power spectra with randomly distributed galaxies and the `gal' dataset are shifted slightly to the left and right, respectively, by 5 per cent on the logarithmic scale. The \modelSA{} model shows strong positive correlations at large scales ($k \lesssim 0.2 ~\mathrm{Mpc}^{-1}$) when correlated with either galaxy dataset, whereas the \modelUni{} model displays negative correlations at these large scales.
    }
    \label{fig:cross_corr_gal}
\end{figure}

We next study the cross power spectrum between \lya{} forest transmission and galaxy distribution. The cross power spectrum $P_{\delta_\mathcal{T},\delta_\mathrm{X}}$ is defined as,
\begin{eqnarray}
    \langle \delta_\mathcal{T}(\pmb{k}) \delta_{X}(\pmb{k'})\rangle = (2\pi)^3P_{\delta_\mathcal{T},\delta_{X}}(k)\delta_\mathrm{D}(\pmb{k}-\pmb{k'}) \ ,
    \label{eq:ps_cross}
\end{eqnarray}
where $\delta_X=X/\Bar{X}-1$, with $X$ representing any field, and $\delta_\mathrm{D}$ is the Dirac delta function. This quantity is the Fourier transform counterpart of the cross-correlation function. Here, $\pmb{k}$ denotes the 3D wave-vector. However, for simplicity, we focus on the spherically averaged case and present $P_{\delta_\mathcal{T},\delta_{X}}$ as a function of the scalar $k$.

In Fig.~\ref{fig:cross_corr_gal}, we show the cross-power spectra of the \lya{} forest transmission maps with the galaxy distribution. This quantity was estimated using Eq.~\ref{eq:ps_cross} by replacing $X$ with ${N_\mathrm{Random}}$, ${N_\mathrm{gal}}$ and ${N_\mathrm{LAE}}$, which are the number of galaxies in the corresponding dataset assigned to a grid of same dimension as the transmission maps. In this figure, we have only considered the reconstructed \lya{} forest transmission. 
A clear distinction emerges between the reionization model containing residual neutral islands (\modelSA{}) and the \modelUni{} model, which lacks such islands. The \modelSA{} model shows positive power at large scales ($k \lesssim 0.2 ~\mathrm{Mpc}^{-1}$), while the \modelUni{} model exhibits negative power. This relation is a strong signature of the presence of neutral islands during the observed epoch. 

\subsection{Cross-correlation with 21-cm signal}\label{sec:result_cross_21cm}

\begin{figure*}
    \centering
    \includegraphics[width=1.0\linewidth]{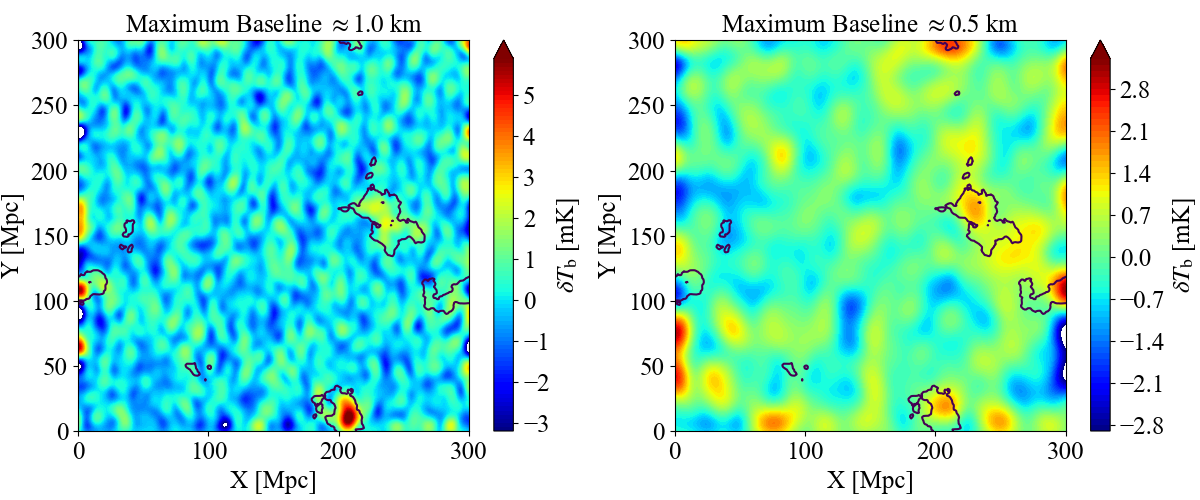}
    \vspace{-0.4cm}  
    \caption{
    21-cm signal slices from the \modelSA{} reionization model, averaged over the same frequency width as the Subaru NB filter used for the \lya{} forest transmission map in this study. These slices include thermal noise corresponding to 3000 (left) and 500 (right) hours of observation with SKA-Low. Neutral islands at simulation resolution are indicated with dark solid contours. 
    \textit{Left panel}: The resolution is degraded assuming a maximum baseline of 1 km, which corresponds to $\sim$12 Mpc at $z=5.7$. This resolution helps \RefereeReport{to} resolve the large neutral islands in the map. 
    \textit{Right panel}: The noise level is higher compared to the left slice, and the image is degraded assuming a maximum baseline of 0.5 km, which corresponds to $\sim$23 Mpc. At this resolution, only the very large neutral islands can be detected.
    }
    \label{fig:Source1_SinkA_21cm_map}
\end{figure*}

\begin{figure}
    \centering
    \includegraphics[width=1.0\linewidth]{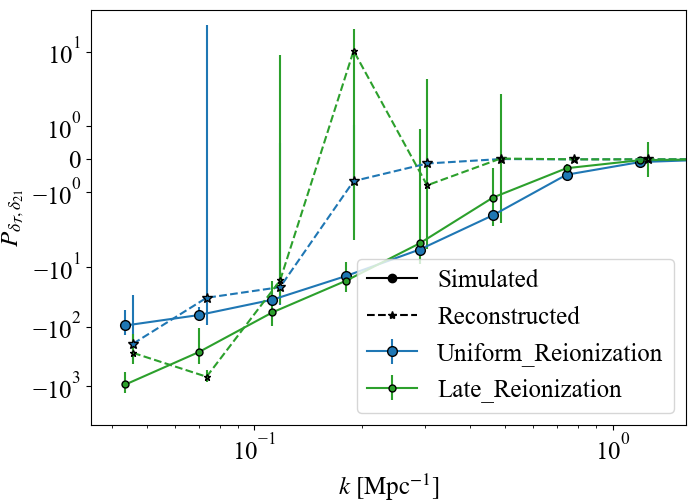}
    \vspace{-0.4cm}  
    \caption{
    Cross-power spectra of the \lya{} forest transmission maps and the 21-cm signal are shown for two reionization models: simulated (filled circles) and reconstructed (filled stars) observations. Error bars correspond to sample variance. The reconstructed data includes the impact of noise (assuming 3000 hours of observation with SKA-Low for the 21-cm signal). For clarity, the power spectra of the reconstructed maps are shifted slightly to the right by 5 per cent on the logarithmic scale. Both models exhibit negative correlations at large scales ($k \lesssim 1 ~\mathrm{Mpc}^{-1}$), a universal feature for any reionization scenario. The cross-power spectra of the reconstructed data approach zero at $k \gtrsim 0.3 ~\mathrm{Mpc}^{-1}$ due to unresolved signal below this scale.
    }
    \label{fig:Source1_SinkA_cross_power}
\end{figure}

\begin{figure*}
    \centering
    \includegraphics[width=1.0\linewidth]{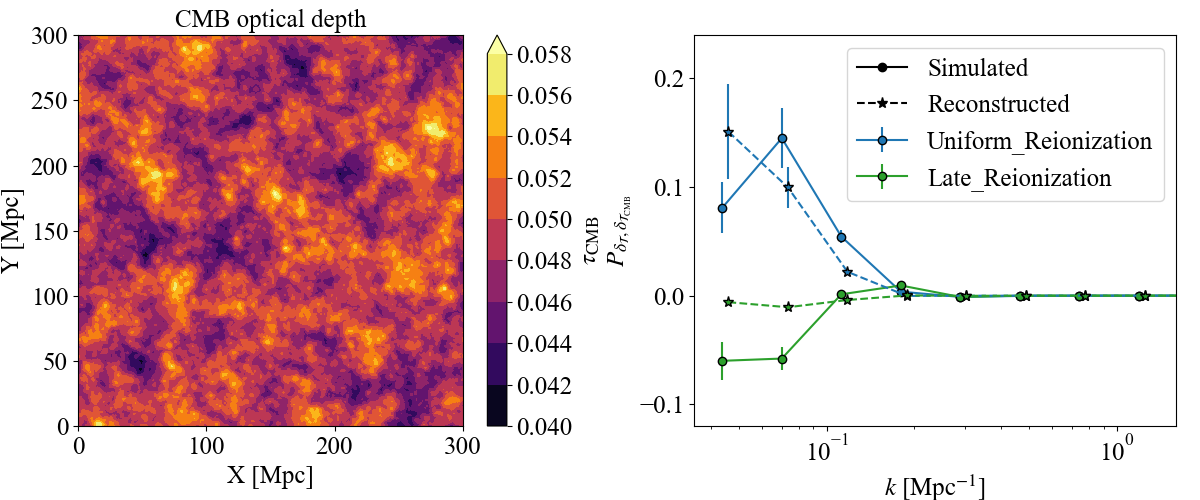}
    \vspace{-0.5cm}  
    \caption{
    The transmission of CMB photons due to Thomson scattering.
    \textit{Left panel}: The patch $\tau_\mathrm{CMB}$ map containing structures that depend on the reionization history along each line-of-sight.
    \textit{Right panel}: The cross-power spectra of the \lya{} forest transmission maps and the CMB photons for the two reionization models are shown. Filled circles and stars represent simulated and reconstructed observations, respectively. For clarity, the data points from the reconstructed maps are shifted to the right by 5 per cent on the logarithmic scale. The \modelUni{} model reveals a positive correlation, while the \modelSA{} model shows negative cross-power at large scales for both simulation and mock observation resolutions.
    }
    \label{fig:cross_corr_CMBtau}
\end{figure*}

In this sub-section, we cross-correlate the \lya{} forest transmission with the 21-cm signal during reionization. This signal is produced by neutral hydrogen and can be measured by radio telescopes as differential brightness temperature that is given as \citep[e.g.,][]{pritchard201221},
\begin{eqnarray}
\label{eq:dTb} &\delta T_{b} (\pmb{r}, z) = T_{0} (z) x_{\rm HI}(\pmb{r},z) \left[ 1+\delta_{b}(\pmb{r},z)\right] \ , \\ 
&T_0 (z) \approx 27 \left(\frac{1+z}{10}\right)^{\frac{1}{2}} \left( \frac{\Omega_\mathrm{b}}{0.044} \frac{h}{0.7} \right) \left( \frac{\Omega_\mathrm{m}}{0.27} \right)^{-\frac{1}{2}} ~\mathrm{mK},
\end{eqnarray}
where $\delta_b$ is the mass density contrast and $x_{\rm HI}$ is the neutral fraction. $T_0$ depends only on the cosmological parameters and the redshift ($z$). 
During the end stages of reionization, the 21-cm signal produced by the neutral hydrogen gas inside galaxies also contributes to the observed signal \citep[e.g.,][]{villaescusa2014modeling,villaescusa2018ingredients,xu2019h}. We included this contribution in addition to the signal from the IGM using the method described in \citet{giri2024end}.

Fig.~\ref{fig:Source1_SinkA_21cm_map} presents slices of the 21-cm signal image from the \modelSA{} simulation, consistent with those shown in previous sections, containing two different noise levels. Although radio telescopes observe the 21-cm signal across multiple frequency channels, we apply a top-hat smoothing in the line-of-sight direction to match the filter width used for modeling \lya{} forest transmission maps. Due to instrumental limitations, realistic 21-cm images from radio telescopes will have lower resolution \citep[see e.g.,][]{giri2018bubble,giri2018optimal}. In this study, we assume the SKA-Low as the observing instrument, which is expected to be powerful enough to produce images \citep{mellema2015hi}. Fig.~\ref{fig:Source1_SinkA_21cm_map} includes thermal noise for observation times of 3000 hours (left panel) and 500 hours (right panel), following the noise modeling scheme of \citet{giri2018optimal}.

In the left panel of Fig.~\ref{fig:Source1_SinkA_21cm_map}, the image is degraded to a resolution corresponding to a maximum baseline of 1 km ($\sim$12 Mpc at $z=5.7$), which approximately matches the resolution of reconstructed \lya{} forest transmission maps with background sources that are brighter than $m_\mathrm{UV}=26.0$. The black contours outline neutral islands in the \modelSA{} simulation. Even though the image is noisy, we find visible structures that correlate with the distribution of the neutral islands. A longer observation time is used here compared to the 1000 hours typically suggested for imaging the reionization-era with the 21-cm signal \citep{giri2018optimal,giri2019neutral}, as the signal is weaker in the final stages of reionization. This extended observation time improves the SNR, enabling the detection of neutral islands. The thermal noise creates bright features in the 21-cm signal image that do not correspond to neutral islands. Several studies \citep[e.g.,][]{giri2019neutral,gazagnes2021inferring,bianco2021deep,bianco2024deep} have developed methods for the automatic detection of these islands in 21-cm signal image data. These methods struggle during these end stages and therefore a cross-correlation with other dataset will be a robust probe of the neutral islands.

The right panel of Fig.~\ref{fig:Source1_SinkA_21cm_map} represents a pessimistic scenario with higher noise (observation time of 500 hours), where the resolution is degraded to a resolution corresponding to a maximum baseline of 0.5 km ($\sim$23 Mpc at $z=5.7$) to maintain a similar SNR as the left panel. Despite this lower resolution, large neutral islands remain discernible. A shorter observation time would further degrade the resolution, preventing the detection of the last remaining neutral islands, which are expected to be $\sim$40 Mpc in scale \citep{giri2024end}. Finally, the noise properties of the 21-cm signal maps differ from those of the reconstructed \lya{} forest transmission maps. Additional challenges include optimal signal extraction in the presence of strong foreground contamination \citep[e.g.,][]{hills2018concerns,kern2021gaussian} and unknown noise sources \citep[e.g.,][]{mevius2022numerical,gan2022statistical}. A joint detection of neutral islands in both datasets would provide a robust confirmation of these structures.


In Fig.~\ref{fig:Source1_SinkA_cross_power}, we present the cross-power spectra of \lya{} forest transmission and the 21-cm signal for the \modelUni{} (blue) and \modelSA{} (green) models. The filled stars represent fields directly from simulations, while the filled circles correspond to modelled mock observations. Both the \lya{} forest transmission and the 21-cm signal during reionization depend on the ionization state of the IGM. While \lya{} forest transmission peaks in ionized regions, the 21-cm signal is strongest in neutral regions. This relation leads to a strong anti-correlation between the two fields, regardless of the reionization model. All cases shown in Fig.~\ref{fig:Source1_SinkA_cross_power} exhibit this negative correlation. Although this metric cannot directly identify neutral islands, it serves as an indicator for detecting the 21-cm signal, which can be challenging to confirm due to factors such as residual foreground signal contamination and radio data calibration errors.

\RefereeReport{We also note in Fig.~\ref{fig:Source1_SinkA_cross_power} that the cross-power spectrum of the mock reconstruction performs worse than the true simulated signal at intermediate scales ($k\approx 0.1-0.3$ Mpc$^{-1}$). This is a direct consequence of the reconstruction process of the \lya{} transmission maps. The Gaussian kernel interpolation used to generate these map effectively smooths over fluctuations on scales comparable to the mean separation of the background galaxies. This process damps the correlated signal at intermediate $k$ values. Additionally, photometric noise in the individual transmission measurements further suppresses the sensitivity at these scales. Together, these effects cause the measured cross-power spectrum to approach zero more rapidly than in the ideal, noise-free case, highlighting that the signal is most robustly detected at the largest scales.}



\subsection{Cross-correlation with CMB polarization measurement}\label{sec:result_cross_CMB}

The density of free electrons is traced by the reionization Thomson scattering optical depth, \(\tau_\mathrm{CMB}\), which affects CMB photons and can be inferred from CMB temperature and polarization data \citep[e.g.,][]{Holder2003opticaldepth,planck2020CMBspectralikelihood}. This parameter quantifies the fraction of CMB photons that have been scattered and screened by free electrons along the line of sight. By analysing global \(\tau_\mathrm{CMB}\), we can gain insight into the reionization history that affects the integrated electron density \citep[e.g.,][]{millea2018cosmic,planck2016constraints,planck2020cosmological}. Furthermore, spatial variations in the optical depth can be reconstructed by examining second-order effects in the CMB \citep[e.g.,][]{holder2007patchytau,dvorkin2009Bmode,dvorkin2009reconstructing,gluscevic2013patchy} and can be used to further enhance our understanding of the reionization history \citep[e.g.,][]{tashiro2013constraining,meerburg2013probing,feng2019searching,ccalicskan2024reconstructing}. This patchy optical depth is expressed as,
\begin{eqnarray}
    \tau_\mathrm{CMB}(\pmb{r}) = \frac{c\sigma_T\Omega_b}{m_P}\int^{z_\mathrm{max}}_0 \frac{x_e(\pmb{r},z)(1+z)^2}{H(z)} dz \ ,
    \label{eq:CMBtau}
\end{eqnarray}
where $c$, $\sigma_T$, $m_P$ and $H(z)$ are the speed of light, Thomson scattering cross-section, the mass of the proton and the Hubble parameter, respectively. $x_e(\pmb{r},z)$ is the distribution of fraction of free electrons at position $\pmb{r}$ and redshift $z$. This patchy $\tau$ field depends on the underlying reionization model and captures its inhomogeneous nature. The upper limit of the integration, $z_\mathrm{max}$, corresponds to a redshift before the start of reionization, as the universe can be assumed to have a uniform neutral fraction from $z_\mathrm{max}$ to the redshift of last scattering.

We present the $\tau_\mathrm{CMB}$ estimated for our \modelSA{} model in the left panel of Fig.~\ref{fig:cross_corr_CMBtau}. To construct this map, we generated a lightcone of the ionization fraction between $5.5<z<21$ by following the method described in \citet{giri2018bubble}. Eq.~\ref{eq:CMBtau} was then solved for this lightcone along the redshift (or line-of-sight) direction. For the \modelUni{}, we assumed the same reionization history at $z>5.7$ and completed reionization afterwards. As the differences between the two models occur during a relatively short period of the reionization epoch, the resulting patchy $\tau_\mathrm{CMB}$ maps appear nearly identical. Visually comparing the structures in these maps with the \lya{} forest transmission maps shown in the right panel of Fig.~\ref{fig:models_tau_map} is not feasible, as $\tau_\mathrm{CMB}$ represents an integrated quantity over the reionization history. Instead, to qualify and quantify the relation between these two fields we cross-correlated these maps, which will be discussed next.

For a consistent comparison, we construct the transmission of CMB photons defined as $\mathcal{T}_\mathrm{CMB} = \exp(-\tau_\mathrm{CMB})$ that we cross-correlate with the \lya{} forest transmission $\mathcal{T}$. We should note that the $\mathcal{T}_\mathrm{CMB}$ is free from observational systematics, as it depends on the method used to recover these maps from future CMB data, which is beyond the scope of this study.
In the right panel of Fig.~\ref{fig:cross_corr_CMBtau}, we present these cross-power spectra for both our simulations (filled stars) and reconstructed maps 
(filled circles). Contrary to the \lya{} forest transmission, the $\tau_\mathrm{CMB}$ encodes the imprint of ionization bubble evolution throughout the entire reionization history. A simple way to interpret the structures in the patchy $\tau_\mathrm{CMB}$ map is that regions of early reionization yield high $\tau_\mathrm{CMB}$ (or low $\mathcal{T}_\mathrm{CMB}$) values, while regions of delayed reionization result in low $\tau_\mathrm{CMB}$ (or high $\mathcal{T}_\mathrm{CMB}$). For example, see fig.~1 in \citet{giri2019neutral} that compared the average $\tau_\mathrm{CMB}$ for different reionization histories. The \lya{} forest transmission is positively correlated with $\mathcal{T}_\mathrm{CMB}$ in the \modelUni{} model but negatively correlated in the \modelSA{} model. In the \modelSA{} model, pixels within the neutral islands correspond to regions with delayed reionization along the line-of-sight. These regions exhibit high $\mathcal{T}_\mathrm{CMB}$ values, which are inversely related to the \lya{} forest transmission, as the latter is suppressed in neutral islands. 
Thus, we conclude that this cross-power spectra can be used to distinguish between reionization models.



\section{Conclusions}\label{sec:conclusion}

In this study, we explore the potential of photometric IGM tomography to map the remaining neutral islands during the final stages of reionization. The \lya{} forest transmission is highly sensitive to the distribution of neutral hydrogen in the IGM, making it a valuable probe of the ionization state. By observing a sufficient number of \lya{} forest transmission sightlines from background galaxies within an observational field, we can effectively reconstruct a detailed map of the ionization state of the IGM.

We model the \lya{} forest transmissions that can be measured with the NB filters available on the Subaru/HSC. To resolve $\sim$8 Mpc 
structures in the IGM in a field of view spanning a few hundred Mpc, 
we find that more than 500 background sources per $\rm (300\,\rm Mpc)^2$ area is required, corresponding to galaxies down to $\sim26$ mag with 20\% spectroscopic confirmation rate. We can reconstruct the ionization state of the IGM with less number of background galaxies, but the resolution will be poor. For example, with $\sim$50 background galaxies per $\rm (300\,\rm Mpc)^2$ equivalent to $\sim24.5$ mag sources, the reconstructed \lya{} forest transmission map has $\sim$45 Mpc resolution. While this can identify rare extremely large neutral islands, it is not ideal to detect more common neutral islands during the end stages of reionization ($z\lesssim 6$). 

The accuracy of reconstructing the \lya{} forest transmission maps significantly depends on the depth of NB imaging, as well as the brightness and number of background galaxies. To quantify the impact of observational errors on identifying neutral islands, we generated a mock tomographic map that accounts for various types of observational noise. The variance map from the mock tomography enables us to assess the effects of observational uncertainty across all analyses. Despite the presence of the observational noise, our findings demonstrate that the neutral islands and various cross-correlation signals remain detectable in the reconstructed \lya{} forest transmission maps.

We analysed structures in the reconstructed \lya{} forest transmission maps to identify meaningful features. To achieve this, we developed an unsupervised feature-finding method by enhancing the thresholding algorithm proposed by \citet{li1993minimum} and \citet{li1998iterative} with the simple linear iterative clustering (SLIC) technique from \citet{achanta2012slic}. This improved method is effective at identifying features in noisy images.
Using this approach, we successfully identified the large-scale distribution of neutral islands. Despite some susceptibility to false positives due to the low signal-to-noise ratio of the reconstructed maps, the island size distribution closely approximated the true distribution. As the volume of data grows substantially in the near future, this analysis demonstrates the potential of image analysis techniques to efficiently process large datasets, extract meaningful subsets, and support further study using robust methods or targeted observations.
Moving forward, we plan to enhance this method by incorporating advanced supervised machine learning algorithms, such as U-Net neural network \citep[e.g.,][]{ronneberger2015unet,siddique2021unet,Gagnon-Hartman:2021erd,bianco2021deep,bianco2024deep,bianco2024serenet}.

Cross-correlation measurements provide a robust strategy to study the large-scale fluctuations of the ionization state of the IGM. 
The average \lya{} forest transmission around galaxies can be used to study how the reconstructed transmission maps are correlated with the galaxy distribution \citep[see][for a detailed discussion]{kakiichi2023photometric}. To understand this correlation, we compare the average \lya{} forest transmission around galaxies with that around randomly distributed points. We find a large-scale excess of \lya{} forest transmission both around LAEs and all galaxies in the \modelSA{} simulation. The cross-correlation signal peaks at approximately 10 Mpc and preferentially decreases at smaller radii close to galaxies. This positive correlation is stronger for LAEs, which are more massive galaxies, consistent with previous studies \citep{garaldi2022thesan,Conaboy2025}. Our mock simulations confirm that the galaxy–\lya{} forest cross-correlation signal can be recovered from photometric IGM tomography. Moreover, in the post-reionized universe—where the IGM is fully ionized, as in the \modelUni{} model—we observe no excess \lya{} forest transmission. Instead, the cross-correlation signal shows a preferential absorption at all scales around galaxies. This supports the idea that the shape of the cross-correlation measured through photometric IGM tomography can provide a precise probe of the final stages of reionization and the disappearance of neutral islands.


We also studied the cross-power spectrum of the \lya{} forest transmission with the galaxy distribution during the end stages of reionization. In the presence of neutral islands, the cross-power spectrum is positive in contrast to the case with no neutral island that shows a negative behaviour. We interpret this as an observational signature of galaxies residing in ionized bubbles \citep[e.g.,][]{barkana2001beginning,iliev2006simulating,iliev2012can,zackrisson2020bubble,bianco2021deep}. 
The cross-power spectrum with massive galaxies or LAEs shows a stronger positive correlation compared to that with the full galaxy population. This signal is consistent with the presence of large-scale neutral islands in the IGM. However, other plausible explanations—such as fluctuations in the ionizing background or variations in the mean free path—may also contribute to or mimic this positive correlation \citep[e.g.,][]{davies2016large,kakiichi2025aspire}. We leave a detailed investigation of these scenarios to future work.

Since \lya{} forest transmission is weak in neutral hydrogen gas, which produces the 21-cm signal, the cross-power spectrum between \lya{} forest transmission and 21-cm signal is consistently negative, regardless of the reionization model. Detecting the 21-cm signal remains challenging due to various contaminants in the data, such as foreground signals, calibration errors, and thermal noise \citep[e.g.,][]{hills2018concerns,mazumder2022observing,gorce2023impact}. While the \lya{} forest–21-cm cross-correlation may not directly detect neutral islands, the observed anti-correlation provides a robust signature confirming the presence of the 21-cm signal.


Lastly, we investigated the cross-correlation between the \lya{} forest transmission and the spatially varying Thomson optical depth or patchy $\tau_\mathrm{CMB}$, which can be measured from CMB data. The patchy $\tau_\mathrm{CMB}$ contains information about the formation and evolution of ionized regions throughout the history of reionization. We \RefereeReport{find} that this field is negatively (positively) correlated with the \lya{} forest transmission in the presence (absence) of neutral islands. This correlation provides an additional method for detecting the last remaining neutral islands.

This study highlights the potential of IGM tomography for investigating neutral islands while also enabling statistical interpretation of the properties of the IGM. In addition to providing summary statistics (e.g., neutral island size distribution, cross-correlation, power spectra, and average transmissions around galaxies) to characterize the end stages of reionization, advanced techniques such as simulation-based inference \citep[SBI; e.g.,][]{cranmer2020frontier} may be employed to extract maximum information from the reconstructed \lya{} forest transmission maps in the future. 

\RefereeReport{
We note a key caveat is that the mass resolution of our dark matter simulation is not ideal for modelling the lowest mass haloes ($\sim 10^9 M_\odot$). However, we have verified that the resulting halo mass function is consistent with theoretical predictions \citep[see Appendix A of][for details]{giri2024end}. A more significant concern is that unresolved sources could fragment the large neutral islands, making them smaller and harder to detect. 
}
\RefereeReportTwo{
While theoretical studies show that star formation is suppressed in these low-mass halos by radiative feedback \citep[e.g.,][]{iliev2005minihalo,nebrin2023starbursts}, it is plausible that this feedback would be ineffective inside the large, self-shielded neutral islands. In this scenario, unresolved low-mass galaxies could form within them and begin to erode the neutral islands. This implies that our specific predictions for island size and detectability may be somewhat optimistic.
However, observational evidence for large neutral islands at these epochs does exist \citep[e.g.,][]{becker2015evidence,kulkarni2019large,keating2020long}, suggesting that large-scale features persist. Therefore, while our model may underestimate the level of fragmentation, the observational strategies developed in this paper remain a valid and necessary approach for finding and characterizing these large-scale structures.
}

In the coming years, several observational programs are expected to significantly advance the capabilities of IGM tomography. Missions such as JWST, Euclid, and the Nancy Grace Roman Space Telescope, along with wide-field spectroscopic surveys using Subaru/PFS and VLT/MOONS, will provide deep and extensive datasets well-suited for tomographic reconstruction. In parallel, Stage-4 CMB experiments will deliver high-quality data capable of mapping the $\tau_\mathrm{CMB}$, enabling joint analyses with IGM tomography to probe the ionization topology during the end stages of reionization with unprecedented precision.


\section*{Acknowledgements}
We acknowledge the Nordita program ``Cosmic Dawn at High Latitudes'' for helping to initiate this project.
Nordita is supported in part by NordForsk. SKG and PDM are supported by NWO grant number OCENW.M.22.307.
KK acknowledges support from VILLUM FONDEN (71574). The Cosmic Dawn Center is funded by the Danish National Research Foundation under grant no. 140.
We acknowledge the allocation of computing resources provided by the National Academic Infrastructure for Supercomputing in Sweden (NAISS) and the Swiss National Supercomputing Centre (CSCS). 
We have utilised the following \texttt{Python} packages for manipulating the simulation outputs and plotting results: {\tt numpy} \citep{van2011numpy}, {\tt scipy} \citep{virtanen2020scipy}, {\tt matplotlib} \citep{hunter2007matplotlib}, {\tt astropy} \citep{robitaille2013astropy} {\tt scikit-image} \citep{scikit-image} and {\tt tools21cm} \citep{giri2020tools21cm}. 

\section*{Data Availability}
The reionization simulation data used in this study are freely available at \url{https://doi.org/10.5281/zenodo.10785609}. 
The source code used for the simulations of this study, the {\tt Pkdgrav3} (\url{https://bitbucket.org/dpotter/pkdgrav3}) and {\tt pyC2Ray}
(\url{https://github.com/cosmic-reionization/pyC2Ray}), are publicly available.
The tools to model \lya{} forest transmission in cosmological reionization simulations are publicly available through {\tt toolslyman} (\url{https://github.com/sambit-giri/toolslyman}).



\bibliographystyle{mnras}
\bibliography{refs} 




\appendix

\section{Calibration of Lyman-$\alpha$ optical depth for low resolution simulation} \label{sec:kres_calibration}

\begin{figure}
    \centering
    \includegraphics[width=1.0\linewidth]{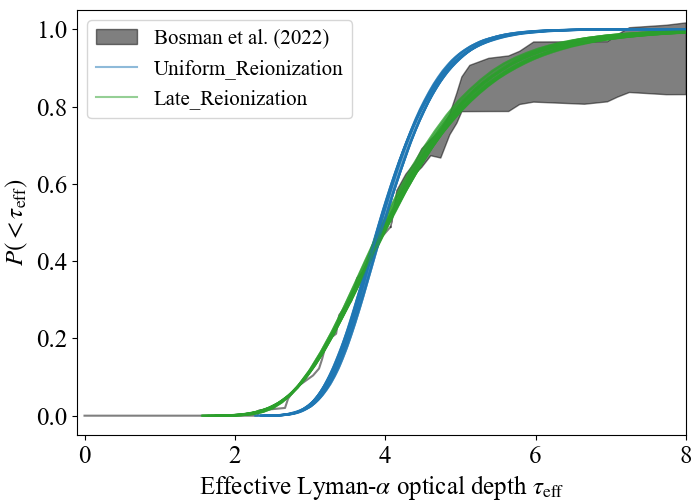}
    \vspace{-0.4cm}  
    \caption{
    Cumulative distribution function (CDF) of the effective \lya{} optical depth ($\tau_\mathrm{eff}$) at $z=5.7$. The grey shaded region correspond to the CDFs estimated from high redshift quasar measurements \citep{bosman2022hydrogen} with $\tau_\mathrm{eff}$ averaged over $\sim$45 Mpc 
    scales. The solid lines give the CDF of the $\tau_\mathrm{eff}$ estimated from the two reionization models used in this work. Multiple curves of each color correspond to random slices drawn from these simulations.
    }
    \label{fig:kres_calibrated_S22}
\end{figure}

The modeling the transfer of \lya{} photons in the intergalactic medium requires high resolution ($\sim$ kilo-parsec scale) simulation of the neutral hydrogen gas in the intergalactic medium (IGM). However, this resolution is computationally expensive for simulations focusing on scales more than about 100 Mpc. 
Therefore, we employed a free parameter $\kappa_\mathrm{res}$ in our modeling of the \lya{} optical depth for our comparatively low resolution cosmological simulations (see Eq.~\ref{eq:tau_i}). 
We calibrated this parameter by comparing it with the \lya{} forest data from \citet{bosman2022hydrogen}.
In Fig.~\ref{fig:kres_calibrated_S22}, we show the cumulative distribution function (CDF) of the effective optical depth ($\tau_\mathrm{eff}$) measured from the data at $z=5.7$, shown as a black shaded region. This $\tau_\mathrm{eff}$ is averaged over 45 Mpc 
scales. 
The blue and green solid curves represent the CDFs estimated from the \modelUni{} ($\kappa_\mathrm{res}=0.16$) and \modelSA{} ($\kappa_\mathrm{res}=0.56$) models, respectively. We used these values consistently throughout the study. 

The curves of the same color represent randomly selected slices from each simulation. Our analysis indicates that the \modelUni{} model fails to reproduce the slope of the observed CDF. In contrast, our realistic \modelSA{} model successfully replicates this slope. It is important to note that the \modelSA{} model was calibrated to the measured ultraviolet background during the final stages of reionization \citep[see][for more details]{giri2024end}. With $\kappa_\mathrm{res}$ parameter, we get similar dynamic range for the \lya{} optical depths in our simulations, allowing consistent comparison.

\section{Structure identification algorithm} \label{sec:struct_ident}

\begin{figure*}
    \centering
    \includegraphics[width=0.95\linewidth]{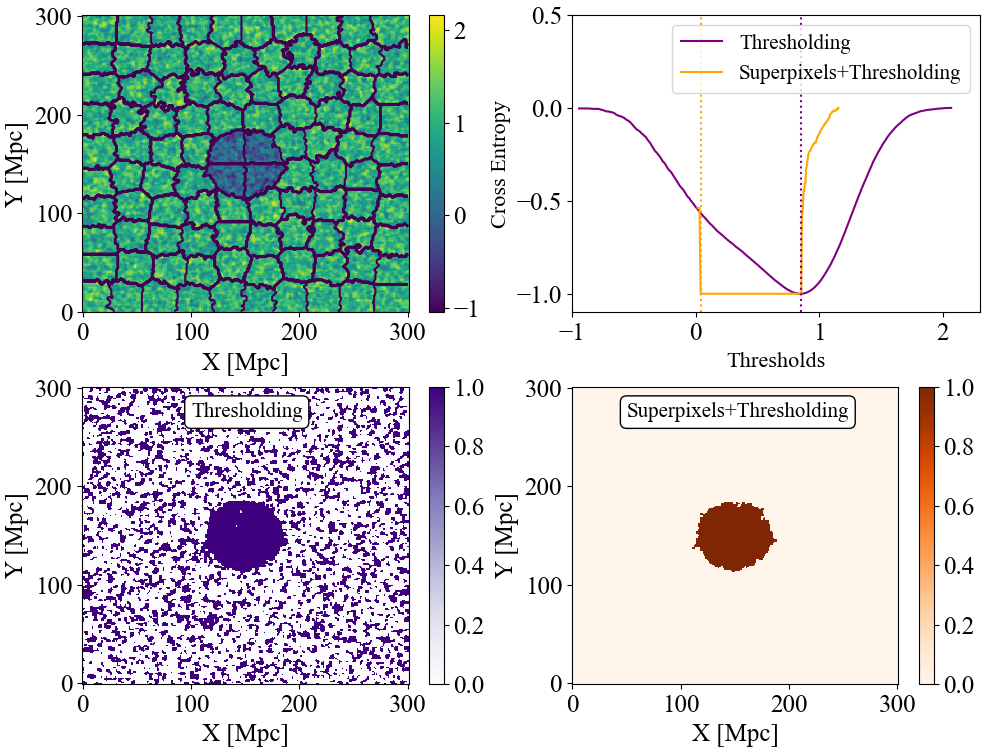}
    \vspace{-0.2cm}  
    \caption{
    An illustration of the structure identification method in noisy images.
    \textit{Top-left panel}: A noisy image with circular structure at the center that we want the method to identify. The dark contours correspond to the superpixels. 
    \textit{Top-right panel}: The cross entropy estimated by applying different threshold values on the true image (purple) and image decomposed into superpixels (orange). These values were normalize to to vary between -1 and 0 for comparison of the two cases. The vertical dotted lines correspond to the thresholds that minimize the cross entropy. 
    \textit{Bottom-left panel}: The binary image showing the structures identified by directly applying the thresholding method of \citet{li1998iterative} on the noisy image in purple. The small regions are erroneous features imprinted by the noise. 
    \textit{Bottom-right panel}: The binary image showing the structures identified by combining the superpixel method with the thresholding method, which better accounts for the noise. 
    }
    \label{fig:toy_sphere_map}
\end{figure*}

We present a modified method for identifying structures in noisy images by combining superpixel techniques with a threshold-based approach. Superpixel algorithms are designed to over-segment images, primarily to detect edges. In the top-left panel of Fig.~\ref{fig:toy_sphere_map}, we show a noisy image containing a circular feature at the center. We apply the Simple Linear Iterative Clustering (SLIC) algorithm \citep[e.g.,][]{achanta2012slic}, using the implementation from the \texttt{scikit-image} Python package \citep{scikit-image}, to segment the image into superpixels—groups of pixels with similar properties. This algorithm has proven effective in similar contexts, such as identifying ionized regions in noisy 21-cm image data \citep{giri2018optimal}. For a broader comparison of superpixel algorithms, we refer readers to \citet{wang2017superpixel}.

The same panel also displays the superpixel boundaries (black contours). Although individual superpixels do not represent complete structures, they outline the shape of the circular feature remarkably well, despite the noise. The number of superpixels, an input parameter for SLIC, is chosen based on the typical scale of meaningful structures in the image. Increasing this number allows for finer segmentation but may also capture more noise. In the example shown, we use 100 superpixels. For the reconstructed \lya{} forest transmission maps, this number is chosen based on the smoothing scale defined in Eq.~\ref{eq:l_smooth}.

To extract continuous structures, we stitch together relevant superpixels using a thresholding technique proposed by \citet{li1998iterative}, which determines an optimal threshold by minimizing the cross-entropy between the foreground and background regions \citep{li1993minimum}. The cross-entropy (CE) is defined as
\begin{eqnarray}
    CE = -\sum^{T}_{i=1}\frac{p_i}{P_\mathrm{fore}}\log\left(\frac{p_i}{P_\mathrm{fore}}\right)-\sum^{N}_{i=T+1}\frac{p_i}{P_\mathrm{back}}\log\left(\frac{p_i}{P_\mathrm{back}}\right) \ ,
\end{eqnarray}
where $p_i$ is the intensity value of the $i^{\rm th}$ pixel. $P_\mathrm{fore}=\sum^{T}_{i=1} p_i$ and $P_\mathrm{back}=\sum^{N}_{i=T+1}p_i$ are the sums over the foreground and background pixels, respectively. $T$ is the number of foreground pixels and $N$ is the total number of pixels in the image. We utilized the implementation of this thresholding method in the \texttt{scipy} Python package \citep{virtanen2020scipy}.

The top-right panel of Fig.~\ref{fig:toy_sphere_map} shows the cross-entropy as a function of the threshold. For comparison, both curves are normalized to range from $-1$ to $0$. The purple curve shows the results of applying thresholding directly to the noisy image, and the corresponding optimal threshold is marked with a vertical line. The binary image created using this threshold (bottom-left panel) contains the circular feature but also includes several noise artifacts.

To mitigate this, we compute a \textit{mean superpixel map} by assigning each pixel the average intensity of its superpixel. This averaging step reduces the impact of noise. The orange curve in the top-right panel shows the cross-entropy values for thresholds applied to this mean superpixel map. Unlike the smooth purple curve, the orange curve appears stepped due to the discretized nature of the superpixels. Any threshold value in the flat minimum of the curve yields a similar result. The binary image obtained using the chosen threshold (orange vertical line) is shown in the bottom-right panel. This output successfully isolates the circular feature while suppressing noise, demonstrating the effectiveness of our modified approach.


\bsp	
\label{lastpage}
\end{document}